\documentclass[11pt,a4paper]{scrartcl}

\usepackage{CLICdp}
\usepackage{CLICdp_definitions}

\usepackage[T1]{fontenc}

\usepackage{hyperref}
\usepackage{url}
\usepackage{mathtools}
\usepackage{wrapfig}
\usepackage{footnote}
\usepackage{subcaption}
\usepackage{caption}
\usepackage{verbatim}
\usepackage{comment}
\usepackage{amssymb,amsmath}
\usepackage{indentfirst}
\usepackage{wasysym}
\usepackage{multirow}
\usepackage{xspace}
\usepackage{acronym}
\usepackage[bottom]{footmisc}

\usepackage[safe]{tipa} 

\newcommand{\secondstage}{1.5\,TeV\xspace}
\newcommand{\thirdstage}{3\,TeV\xspace}
\renewcommand{\epem}{$e^{+}e^{-}$\xspace}
\newcommand{\delphes}{\textsc{Delphes}\xspace}
\newcommand{\ggtohad}{$\gamma \gamma \to$ had.\xspace}
\newcommand{\chmassdiff}{$m_{H^{\pm}} - m_{H}$\xspace}
\newcommand{\hphm}{H$^+$H$^-$\xspace}

\title{Pair-production of the charged IDM scalars \\ at high energy CLIC}

\addauthor{Jan~Klamka}{}
\addauthor{Aleksander~Filip~\.Zarnecki}{}

\addinstitute{0}{Faculty of Physics, University of Warsaw, Poland}

\date{Revision for EPJ C\\ \today}
\clicdppub{2022}{001}  

\titlecomment{This work was carried out in the framework of the CLICdp
              Collaboration} 

\abstract{
    The Inert Doublet Model (IDM) is a simple extension of the
    Standard Model, introducing an additional Higgs doublet that
    brings in four new scalar particles. The lightest of the IDM
    scalars is stable and is a good candidate for a dark matter
    particle.
    The potential of discovering the IDM scalars in the experiment at
    the Compact Linear Collider (CLIC), an $e^{+}e^{-}$ collider proposed
    as the next generation infrastructure at CERN, has been tested for
    two high-energy running stages, at \secondstage{} and
    \thirdstage{} centre-of-mass energy.
    The CLIC sensitivity to pair-production of the charged IDM scalars
    was studied using the full detector simulation with \geant for
    selected high-mass IDM benchmark scenarios and the semi-leptonic
    final state.
    To extrapolate full simulation results to a wider range of IDM
    benchmark scenarios, the CLIC detector model defined in the \delphes
    fast simulation framework was modified to take into account the
    \ggtohad beam-induced background. 
    Results of the study indicate that heavy charged IDM scalars can
    be discovered at CLIC for most of the considered benchmark
    scenarios, up to masses of the order of 1\,TeV.
}

\graphicspath{ {./plots/} }

\begin{document}

\titlepage


\section{Introduction}

Since the 2012 discovery of a particle with properties of the Standard
Model (SM) Higgs boson~\cite{Chatrchyan:2012ufa,Aad:2012tfa}, no
direct 
observation of any new physics, not described within the SM,
has been made at the Large Hadron Collider (LHC), nor in any other high 
energy physics experiment. 
Still, many indications suggest the existence of physics Beyond 
the Standard Model (BSM).
This includes in particular numerous astrophysical and cosmological observations pointing to
the existence of dark matter (DM) in our Universe, constituting to about 27\% of the total energy budget \cite{Planck:2018vyg}.
The SM alone is not able to explain the fact that neutrinos have mass, deviations from the SM predictions are observed in precision low-energy observables such as 
$g_\mu -2$ \cite{Muong-2:2021ojo} and some B meson decays \cite{LHCb:2021lvy,LHCb:2021trn}.
Also the recent high-precision measurement of the W boson mass with the CDF II detector \cite{CDF:2022hxs} is in significant tension with the SM expectation.

The scalar sector of the SM leaves lots of room for new physics and its extensions are considered in a wide range of BSM models (see eg. \cite{Robens:2022zav} and references therein). At the same time it is still the least tested part of the theory.
So far only the mass of the new particle has been determined with high precision 
\cite{ATLAS:2015yey,ATLAS:2018tdk,CMS:2020xrn}.
While the Higgs boson couplings to other SM particles are being
tested with increasing precision~\cite{Sirunyan:2018koj,ATLAS:2019nkf,CMS:2020xwi},
its self-coupling, needed to confirm the shape of the scalar
potential and fully validate the model, is basically not constrained.
This will be a very challenging task
for the LHC and its planned upgrade to higher luminosity, but also
for the proposed future colliders~\cite{deBlas:2019rxi}. 

A consensus has formed in recent years
in the particle physics community, confirmed by the 2020 Update of
the European Strategy for Particle Physics \cite{CERN-ESU-015}, that  an electron-positron Higgs factory is the
highest-priority next collider. Within the variety of
proposals~\cite{bambade2019international,Charles:2018vfv,Abada:2019zxq,CEPCStudyGroup:2018rmc,Bai:2021rdg},
the Compact Linear Collider (CLIC) at CERN~\cite{Charles:2018vfv} 
gives the best prospects for the direct BSM searches at the energy frontier.
CLIC is planned to be built in three energy stages,
starting from 380\,GeV, extending to 1.5\,TeV and finally to 3\,TeV,
with corresponding total integrated luminosities of 1000\,\fbinv,
2500\,\fbinv and 5000\,\fbinv, respectively.
The first running stage, focused on the Higgs boson measurements
(Higgs factory), will also allow for the precision measurements of top quark
pair-production (including the threshold scan).
The subsequent stages will mainly focus on direct searches for
BSM physics, but additional Higgs and top-quark measurements
  will also become possible.
A dedicated detector concept for CLIC (CLICdet) was proposed,
optimised for the concept of Particle Flow analysis at TeV energies,
as well as for the treatment of beam-induced backgrounds expected at
CLIC due to high bunch rate and beam intensity \cite{Arominski:2018uuz}. 

In this paper we investigate the CLIC potential for discovering new heavy
scalar particles pair-produced in \epem collisions. We consider the Inert Doublet Model (IDM), a simple extension of
the SM, introducing four new scalars with the lightest
neutral one being a good candidate for a dark matter (DM) particle.
This is a challenging scenario, as two DM particles are always produced and the observed SM final state can be very soft.
It  was already considered in~\cite{Kalinowski:2018kdn}, looking at events with two leptons in
the final state. The study was performed on the generator level,
including the CLIC luminosity spectra and cuts reflecting the expected
detector acceptance. 
A leptonic signature allows for very efficient background supression.
However, the CLIC sensitivity to IDM scalar pair-production in  this
channel is limited to masses below $\sim$500\,GeV due to the small
branching ratio and to the fact that the production cross section decreases with scalar
masses and collision energy. 

Considered in this paper is the charged IDM scalar pair-production
with semi-leptonic final states, as the expected event rates are  
order of magnitude higher than those for for the leptonic final states. 
The analysis is based on the CLICdet model and the \geant-based
simulation tools to simulate the detector response for a limited number
of IDM benchmark scenarios, and is extended to other benchmark scenarios with
a fast simulation technique based on \delphes, as described in
Sec.~\ref{sec:sim}. 
Details of the event selection procedure are presented in
Secs.~\ref{sec:presel} and \ref{sec:mva}, and final results are
discussed in Sec.~\ref{sec:results}.
When describing details of the analysis we focus mainly on the full
detector simulation results.
The same analysis approach has also been used for fast simulation results,
with small modifications which are described in detail.


\section{Inert Doublet Model}

The Inert Doublet Model (IDM) \cite{Deshpande:1977rw, Ilnicka:2015jba}
extends the SM by only one additional doublet in the scalar sector, making it one of the simplest extensions of the SM. The scalar
potential in this model contains the SM-like Higgs doublet,
$\phi_{S}$, and the so-called inert (or dark) doublet, $\phi_{D}$,
that contains four new scalar fields: H$^{\pm}$, A and H. Due to the
additional $Z_{2}$ symmetry, under which the inert doublet is odd, the
new scalars do not interact with SM fermions (on tree-level) and the
lightest of the IDM scalars (H) is stable, hence it is a good DM candidate. 
    
After electroweak symmetry breaking, the model contains seven free
parameters. Fixing the SM-like Higgs boson mass $m_{h}$ and 
the Higgs field vacuum expectation value \textscriptv\xspace to
the SM values, the following set of physical parameters is selected
\cite{Ilnicka:2015jba}: three scalar masses, $m_{H^{\pm}}$, $m_{A}$
and $m_{H}$, and two couplings, $\lambda_{2}$ and $\lambda_{345}$.%
\footnote{For more information about the model, please refer to
\cite{Ilnicka:2015jba,Kalinowski:2018ylg} and references therein.} 
Two sets of IDM benchmark points were proposed
in~\cite{Kalinowski:2018ylg}, based on the scan over the whole
5-dimensional IDM parameter space, taking into account all existing
theoretical and experimental constraints. The points were selected to
cover all interesting areas of parameter space and to respect 
a wide range of dark scalar masses and mass splittings.

In this paper, 23 out of 41 benchmark points presented in
\cite{Kalinowski:2018ylg} are considered:
all high-mass benchmark points (HP) and three
low-mass scenarios (BP) with the highest scalar masses:
BP18, BP21 and BP23. All benchmark points considered in the study,
together with associated model parameters, are shown in 
Table~\ref{tab:bmhigh}. 
More information about the benchmarks, as well as the exact
constraints on the model parameters, can be found in
\cite{Kalinowski:2018ylg}.

\begin{table}[htp]
        \begin{center}
        	\hspace*{3 cm}
            \caption{Benchmark points considered in study, accessible at  $e^+e^-$ colliders with $\mathcal{O}\left( \ensuremath\mathrm{TeV}\right)$ center-of-mass energies. $M_h=125.1\,\ensuremath\mathrm{GeV}$ for all points. BP21 and HP10 provide 100\% dark matter relic density \cite{Kalinowski:2018ylg}. \label{tab:bmhigh} }
            \begingroup
                \small
                \renewcommand*{\arraystretch}{1.3}
            	\setlength{\tabcolsep}{4pt}
                \begin{tabular}{|l|l|l|l|c|c|c|l|l|l|}
\hline 
\multirow{2}{*}{No.} & \multirow{2}{*}{$M_H$ [GeV]} & \multirow{2}{*}{$M_A$ [GeV]} & \multirow{2}{*}{$M_{H^\pm}$ [GeV]} & $Z$ & $W$ & DM & \multirow{2}{*}{$\lambda_2$} & \multirow{2}{*}{$\lambda_{345}$} & \multirow{2}{*}{$\Omega_H h^2$}\\[-1mm] 
 & & & & on-shell &  on-shell & $>$50\%  &  &  & \\ 
\hline 
BP23 & 62.69 & 162.397 & 190.822 &  $\checked$&$\checked$&$\checked$& 2.63894 & 0.0056 & 0.064038\\ \hline
BP18 & 147 & 194.647 & 197.403 &&&& 0.387 & -0.018 & 0.0017718\\ \hline
\textbf{BP21} & 57.475 & 288.031 & 299.536 & $\checked$&$\checked$&$\checked$& 0.929911 & 0.00192 & 0.11946\\ \hline
HP1 & 176 & 291.36 & 311.96 & $\checked$ & $\checked$ &           &  1.4895 & -0.1035 & 0.00072156 \\ \hline 
HP2 & 557 & 562.316 & 565.417 &           &           & $\checked$  &  4.0455 & -0.1385 & 0.072092 \\ \hline 
HP3 & 560 & 616.32 & 633.48 &           &           &           &  3.3795 & -0.0895 & 0.001129 \\ \hline 
HP4 & 571 & 676.534 & 682.54 & $\checked$ & $\checked$ &           &  1.98 & -0.471 & 0.00056347 \\ \hline 
HP5 & 671 & 688.108 & 688.437 &           &           &           &  1.377 & -0.1455 & 0.024471 \\ \hline 
HP6 & 713 & 716.444 & 723.045 &           &           &           &  2.88 & 0.2885 & 0.035152 \\ \hline 
HP7 & 807 & 813.369 & 818.001 &           &           &           &  3.6675 & 0.299 & 0.032393 \\ \hline 
HP8 & 933 & 939.968 & 943.787 &           &           & $\checked$  &  2.9745 & -0.2435 & 0.09639 \\ \hline 
HP9 & 935 & 986.22 & 987.975 &           &           &           &  2.484 & -0.5795 & 0.0027958 \\ \hline 
\textbf{HP10} & 990 & 992.36 & 998.12 &           &           & $\checked$  &  3.3345 & -0.051 & 0.12478 \\ \hline 
HP11 & 250.5 & 265.49 & 287.226 &           &           &           &  3.90814 & -0.150071 & 0.00535 \\ \hline 
HP12 & 286.05 & 294.617 & 332.457 &           &           &           &  3.29239 & 0.112124 & 0.00277 \\ \hline 
HP13 & 336 & 353.264 & 360.568 &           &           &           &  2.48814 & -0.106372 & 0.00937 \\ \hline 
HP14 & 326.55 & 331.938 & 381.773 &           &           &           &  0.0251327 & -0.0626727 & 0.00356 \\ \hline 
HP15 & 357.6 & 399.998 & 402.568 &           &           &           &  2.06088 & -0.237469 & 0.00346 \\ \hline 
HP16 & 387.75 & 406.118 & 413.464 &           &           &           &  0.816814 & -0.208336 & 0.0116 \\ \hline 
HP17 & 430.95 & 433.226 & 440.624 &           &           &           &  3.00336 & 0.082991 & 0.0327 \\ \hline 
HP18 & 428.25 & 453.979 & 459.696 &           &           &           &  3.87044 & -0.281168 & 0.00858 \\ \hline 
HP19 & 467.85 & 488.604 & 492.329 &           &           &           &  4.12177 & -0.252036 & 0.0139 \\ \hline 
HP20 & 505.2 & 516.58 & 543.794 &           &           &           &  2.53841 & -0.354 & 0.00887 \\ \hline 
\end{tabular}

            \endgroup
        \end{center}
    \end{table}

Production of IDM scalars at lepton colliders is dominated by 
production of neutral or charged scalar pairs via couplings of dark
scalars to SM gauge bosons: 
    \begin{eqnarray*}
             e^{+}e^{-} &\to & \mathrm{H}\;\;\mathrm{A}\;, \\ 
             e^{+}e^{-} &\to & \mathrm{H}^{+}\mathrm{H}^{-}.
    \end{eqnarray*}
For the neutral scalar pair production, the produced dark scalar A
decays to a (real or virtual) Z boson and the (lighter) neutral
scalar H, $\mathrm{A} \to \mathrm{Z} \mathrm{H}$, while the produced charged boson H$^\pm$ decays
predominantly to a (real or virtual) W$^\pm$ boson and the neutral
scalar H, $\mathrm{H}^\pm \to \mathrm{W}^\pm  \mathrm{H}$.
Decays involving SM states only are forbidden due to the $Z_{2}$ symmetry.
The two possible production channels can be thus written as:
    \begin{eqnarray}
        e^{+}e^{-} &\to & \mathrm{H}\;\;\; \mathrm{A} \;\;\; \to
        \mathrm{H} \; \; \mathrm{H}\;\;\mathrm{Z}\;, \nonumber \\
        e^{+}e^{-} &\to & \mathrm{H}^{+}\mathrm{H}^{-} \to \mathrm{W}^{+}\mathrm{W}^{-}\mathrm{H}\;\;\mathrm{H}\;.
    \end{eqnarray}
The sensitivity of CLIC to neutral and charged IDM scalar pair-production,
for leptonic decays of the produced $Z$ and $W^\pm$ bosons, was
studied in details in \cite{Kalinowski:2018kdn}.
For CLIC running at 380\,GeV, discovery of the IDM scalars is possible
for most of the benchmark scenarios where dark scalar production is
kinematically allowed at this stage, for $m_A+m_H<290$\,GeV and
$2m_{H^\pm}<310$\,GeV.   
However, for CLIC running at higher centre-of-mass energies, at
1.5\,TeV and 3\,TeV, the discovery-reach increases only to about
500\,GeV, as it is limited by the production cross section decreasing
fast with collision energy.

\begin{figure}[tb]
\includegraphics[width=0.49\textwidth]{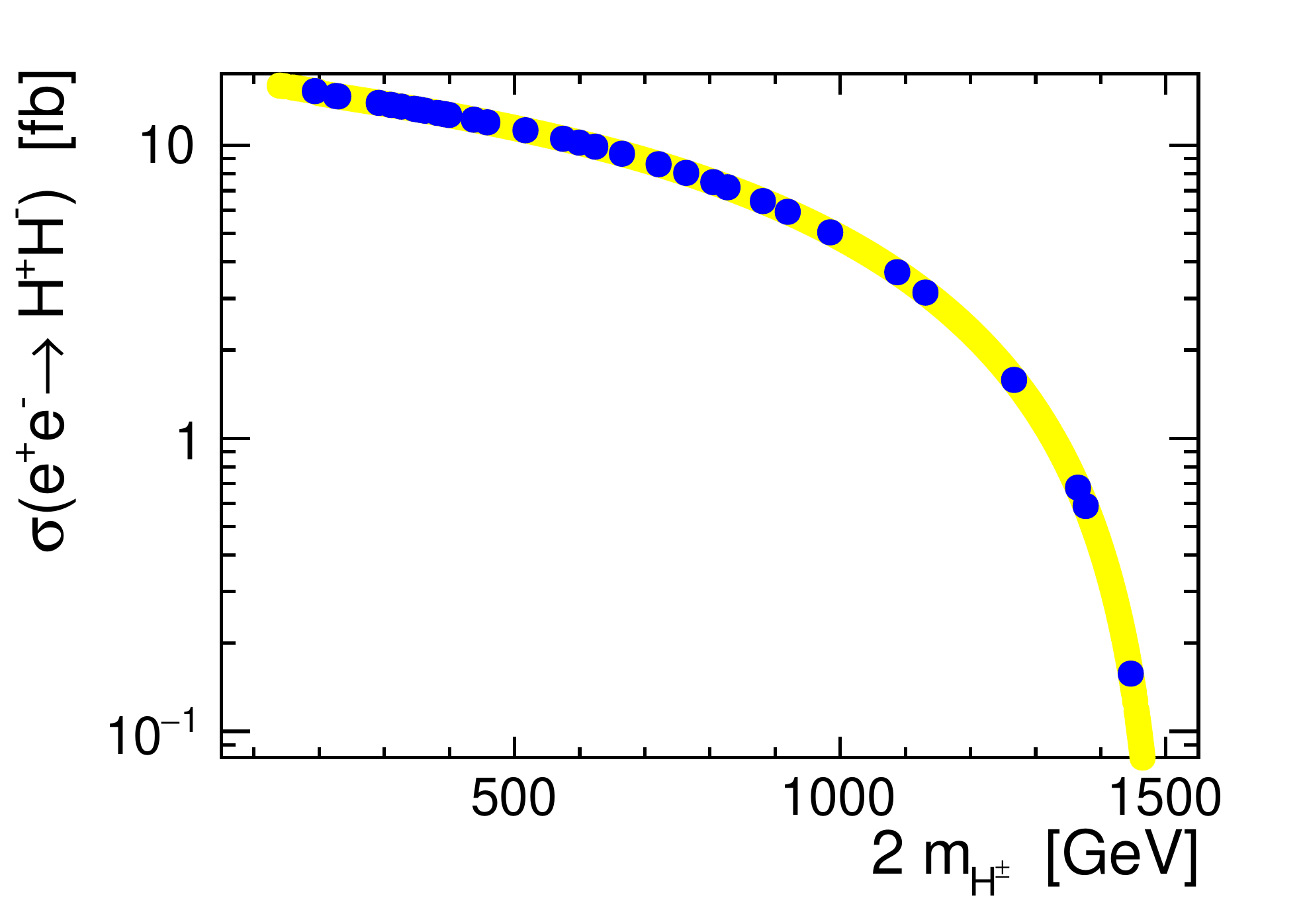}
\includegraphics[width=0.49\textwidth]{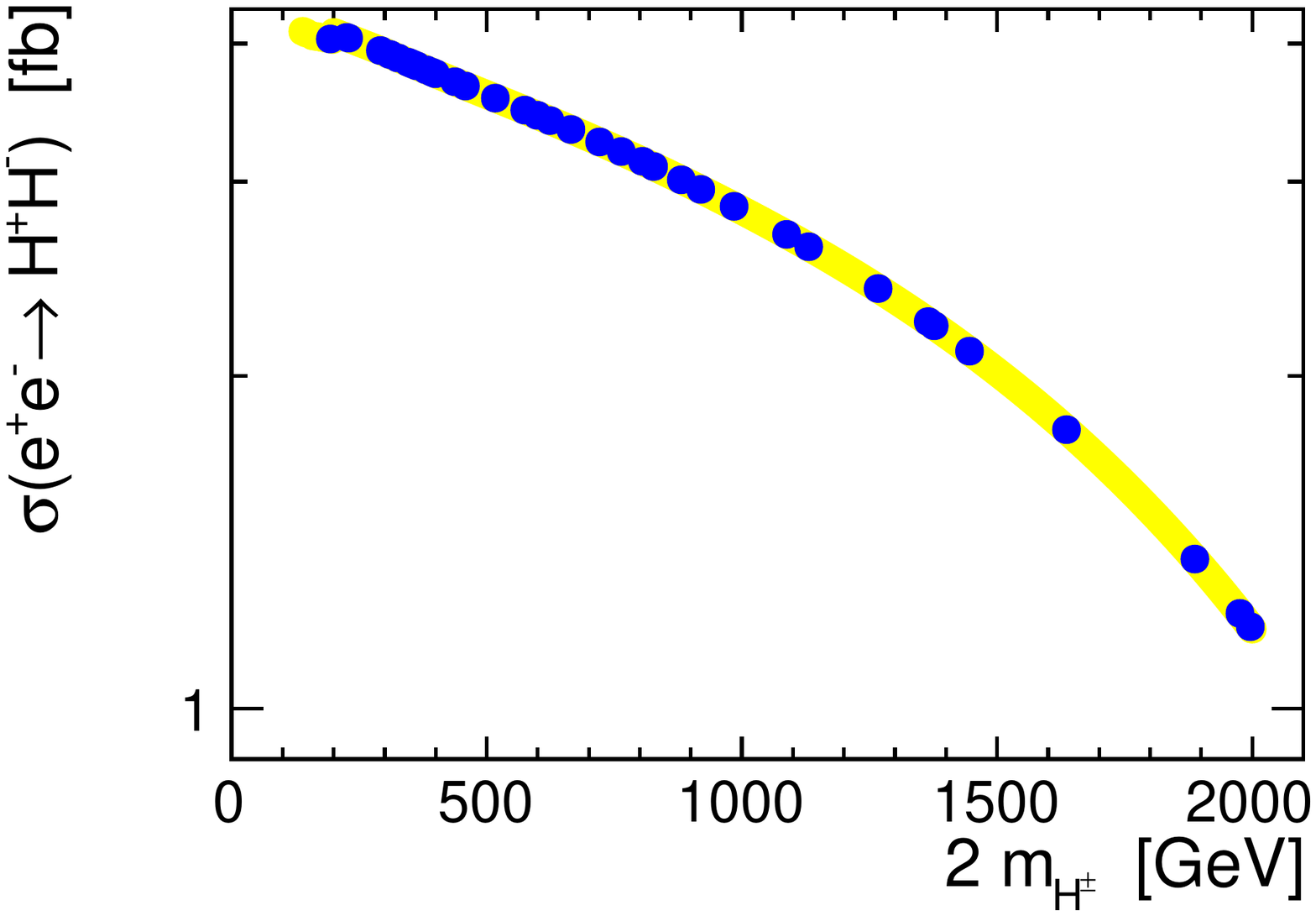}\\[-1.73cm]
                {\hspace*{0.1\textwidth}\scriptsize $\sqrt{s}=$1.5 TeV}\\[-0.53cm]
                {\scriptsize \hspace*{0.59\textwidth} $\sqrt{s}=$3 TeV}\\[0.5cm]
\caption{ Leading-order cross sections for the charged inert scalar
  production, $e^+e^-\to \mathrm{H}^+\mathrm{H}^-$, for collision
  energy of 1.5\,TeV (left) and 3\,TeV (right). The yellow band
  represents all scenarios selected in the model
  scan~\cite{Kalinowski:2018ylg} while the blue dots represent the
  selected benchmark scenarios. Beam energy spectra are not included.
  Figure taken from \cite{Zarnecki:2020swm}.}\label{lo_cros} 
\end{figure}

Considered in the presented study is the \hphm production at high energy
stages of CLIC with the semi-leptonic final state, offering higher decay
rates and hence also higher statistics.
Gauge bosons produced in the final state can be both on-
or off-shell, depending on the dark scalar mass difference,  \chmassdiff.
Reconstruction of the invariant mass of the hadronicaly
decaying W$^\pm$ allows also for better background suppression
for events with virtual W$^\pm$ production.
The cross sections for the \hphm production at 1.5\,TeV and 3\,TeV
CLIC are presented in Figure~\ref{lo_cros}. They depend mostly on the
scalar masses; the influence of couplings $\lambda_2$ and $\lambda_{345}$ is
marginal.


\section{Signal and background simulation \label{sec:sim}}

The CLIC potential for the discovery of charged IDM scalars was studied using
two complementary approaches. First, five selected benchmark scenarios
were studied with full detector response simulation based on \geant.
Then, using the full simulation results to verify and tune the fast simulation
model, the analysis was extended to the set of 23 benchmark points to
estimate the CLIC sensitivity also in other parts of the parameter space.
Only the semi-leptonic final state was considered in generating the signal samples.

\subsection*{\centering Full detector description}

Fully simulated event samples were generated with \textsc{Whizard
  2.7.0}~\cite{Kilian:2007gr}, and 
\textsc{Pythia6}~\cite{Sjostrand:2014zea} was used for
hadronisation. The Monte Carlo particles were propagated through the
\textsc{Geant4}~\cite{AGOSTINELLI2003250} detector model for CLICdet~\cite{Arominski:2018uuz} 
(\textit{CLIC\_o3\_v14}), described by the DD4Hep
toolkit~\cite{Frank_2014}. The reconstruction of physical objects in
the detector was based on \pandora~\cite{Thomson_2009,Marshall:2015rfa}.
The \textsc{iLCDirac} interface~\cite{ilcDiracLC} was used
for the job handling and submission to the grid resources.

Due to high beam intensities and small intervals between subsequent bunches,
beam-induced backgrounds are important and have to be properly
taken into account in the event reconstruction at CLIC.
In the context of the detector performance, the most important 
contribution comes from the photons radiated due to the beam-beam interactions.
Hadrons produced in soft $\gamma \gamma$ collisions
\textit{overlay} on the \epem events  significantly bias 
object reconstruction.
These effects are taken into account in the full simulation, together
with a reconstruction procedure aimed at reducing the impact 
of this background. This turns out to be crucial for this particular
analysis, taking into account that in part of the IDM scenarios the gauge
boson is produced off-shell. Its low mass results in low energy and
momenta of its decay products, what makes their reconstruction more
vulnerable to the influence of soft particles from the overlay events. 

To model the real beam interactions at CLIC, each full simulation
event includes 30 bunch crossings (BX) on top of the physical
event produced in hard interaction, 10 before and 20 after it (the physical
event is placed in the 11th BX).
For each of the additional BX,  on average 1.3 and 3.2
$\gamma \gamma \to$ hadrons events are overlaid at \secondstage and \thirdstage,
respectively.
To suppress this background, dedicated cuts on the timing of
detector hits and reconstructed objects are applied.
First, only hits from 10\,ns following the physical event
are accepted for event reconstruction (which roughly corresponds to 20
BX for 0.5\,ns bunch separation at CLIC) \cite{Arominski:2018uuz}.
Subsequently, cuts are applied to the reconstructed
Particle Flow Objects (PFOs), as described in the CLIC Conceptual Design
Report (CDR)~\cite{clic_cdr}.
The cuts, defined separately for 1.5\,TeV and 3\,TeV CLIC running,
depend on the reconstructed object type (track, neutral hadron or
photon), transverse momentum and the polar angle.
Reconstructed PFOs are accepted for final event reconstruction, if
their reconstructed arrival time differs from a nominal time of a
physical event by less than the corresponding time threshold $t_{\mathrm{cut}}$. 

Particle flow approach assumes reconstruction and identification, by combining information from all subdetectors, of each visible particle in an event. The high calorimeter granularity is essential for this goal. 
Full simulation studies \cite{Arominski:2018uuz} indicate that, even for complex events, muons above 10\,GeV can be identified with more than 98\% efficiency for all energies and polar angles, while electrons are correctly identified in 85\% to 90\% of cases at energies of 20 GeV and higher.
After the reconstruction and selection of PFOs in an event, isolated
leptons (electrons and muons) were selected with the
IsolatedLeptonFinder \marlin processor \cite{marlinleptonfinder},
using the so-called polynomial isolation criterion.
For candidate PFOs with energy above the minimum value of 5\,GeV,
the cut on the energy in a cone surrounding the track was applied, given by
the polynomial dependence on the track energy~\cite{Klamka:2728552}.
We also used the so-called lepton dressing option of the
IsolatedLeptonFinder, correcting for possible electron bremsstrahlung
in the tracking detectors by merging close-by photons and electrons.
For the considered charged IDM scalar pair-production with the semi-leptonic final state, W$^\pm$ decays into tau leptons were also included.
However, identification of hadronic tau decays (tau-jet tagging) was not used and only events with isolated electron or muon were accepted. 
This results in about 20\% loss of signal selection efficiency.
The final efficiency for isolated lepton identification for signal events, for the considered semi-leptonic final state, is about 65--70\%. 
It decreases to about 50--60\% for scenarios with the lowest scalar mass difference.
The purity of the isolated lepton selection (probability of the selected PFO matching closely the generator level lepton) is about 97--98\% for 1.5\,TeV CLIC  and about 94--96\% for 3\,TeV running.\footnote{Lepton identification purity is mainly affected by signal events with hadronic tau decays. Purity of up to 99\% is obtained when only electron and muon decays of W boson are considered.}

PFOs not classified as isolated leptons
were further used for jet clustering. Jets were reconstructed in the
exclusive mode, with two jets in the final state, using the Valencia
Linear Collider (VLC) algorithm~\cite{Boronat:2016tgd}.
This algorithm was designed to be least sensitive to the influence of
the overlay events, and is assumed to be the best choice for the jet
reconstruction at CLIC high-energy stages.
Parameters of the algorithm were set to $\gamma=\beta=1$,
$R=0.9$ for 1.5\,TeV CLIC running and $R=1.2$ for 3\,TeV. The choice of
the $R$ parameter was based on the shape and placement of the peak
corresponding to an on-shell W$^\pm$ boson.
Isolated photons were also identified. A photon with transverse
momentum $p_{T}^{\gamma}$ was considered isolated, if
$p_{T}^{cone}/p_{T}^{\gamma} \leq 0.2$, where $p_{T}^{cone}$ is the
total transverse momentum of other particles in a cone, which is
defined by the squared radius $(\Delta R)^{2} = (\Delta \theta)^{2} +
\sin^{2}(\theta) \cdot (\Delta \phi)^{2} < 0.1$, 
surrounding photon direction.

\subsection*{\centering Fast detector simulation \label{subsec:dataproc_fastsim}}

To extend the study to a larger number of benchmark scenarios, the realistic fast
simulation toolkit \delphes~\cite{deFavereau:2013fsa} was used, version 3.4.2,
with CLICdet detector model cards \cite{Leogrande:2019qbe}, based on
the full simulation results~\cite{Weber:2648827}.
The object reconstruction implemented in the \delphes model follows the
Particle Flow approach of the full simulation reconstruction.
However, particle identification is based on the true MC information,
taking into account finite identification efficiency only.
Isolation criteria are also simplified compared to the full reconstruction:
an electron, muon or photon was considered isolated, if the total $p_{T}$ of
other particles in the cone of radius $\Delta R=0.5$ surrounding it
was less than 0.12 of the candidate particle $p_{T}$.\footnote{These isolation criteria differ from the default CLICdet cards settings described in~\cite{Leogrande:2019qbe}. }
The jets were reconstructed using VLC algorithm in the exclusive mode
again, with the parameters $\gamma=\beta=1$ and $R=1.0$ ($R=1.2$) for
1.5\,TeV (3\,TeV).
As for the full detector simulation, selected values of R
correspond to the best reconstruction of the W$^\pm$ peak.

As a part of the presented study,
the CLICdet cards for \delphes were also modified to take into account
the beam-induced background.
The \textsc{PileUpMerger} module of \delphes, designed to include pile-up
events in hadron colliders, was used to add \ggtohad events to the event record.
Overlay background events generated with \textsc{Pythia6} cannot be used directly, as timing cuts used to select reconstructed PFOs
are not implemented in the CLICdet model.
Hence, a dedicated pre-processing of the overlay events was implemented to take into account
the influence of these cuts  and make the evaluation of
the CLIC sensitivity to IDM scalars more realistic. 

The average number of \ggtohad events added to each physical event is
$20\cdot n_{had}$, where 20 comes from the number of BX that survive
the primary 10\,ns window cut and $n_{had}$ is the expected number of
\ggtohad events in a single BX (at the given CLIC energy stage).
This gives on average 26 background events at \secondstage and
64 at \thirdstage \cite{clic_cdr}, where the number of events is drawn
from the Poisson distribution. 
To model the PFO timing cuts used in the full reconstruction,
particles in an overlay event-file are pre-selected based on their
event number. 
Considering that the BX separation at CLIC is 0.5\,ns, timing cut of
$t_\mathrm{cut}$ applied to a given particle category (particle type and
transverse momentum range) is modelled by accepting particles from
the first $N_{BX} = t_{\mathrm{cut}}/0.5$\,ns events out of every 20
events.
While neglecting the time resolution, this procedure assures that the
number and proportions of particles of different types and kinematic
properties passing the timing cuts are preserved with respect to the
full simulation.   
However, the possible impact of timing cuts on the reconstruction of particles
coming from the physical events is not taken into account.
Also, particles from the overlay events are rejected before the PFOs
reconstruction, so possible effects due to spacial overlap of detector deposits are neglected.

\begin{figure}[tb]
	    \centering
	 	 \begin{subfigure}{.5\textwidth}
	 	 	\centering
	 	 	\includegraphics[width=\linewidth]{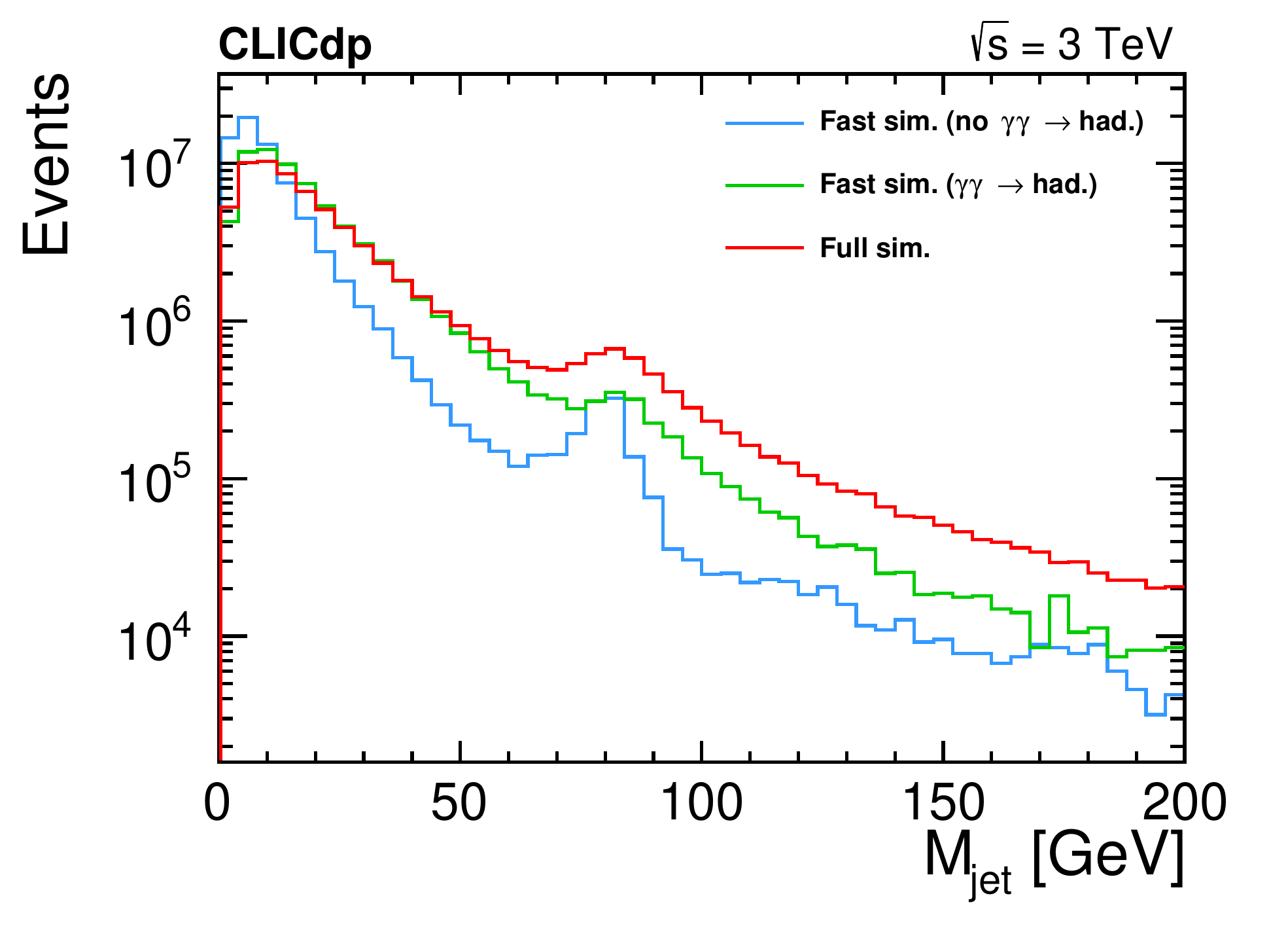}
	 	 	\label{mjj1500overlay}
	 	 \end{subfigure}%
	 	 \begin{subfigure}{.5\textwidth}
	 	 	\centering
	 	 	\includegraphics[width=\linewidth]{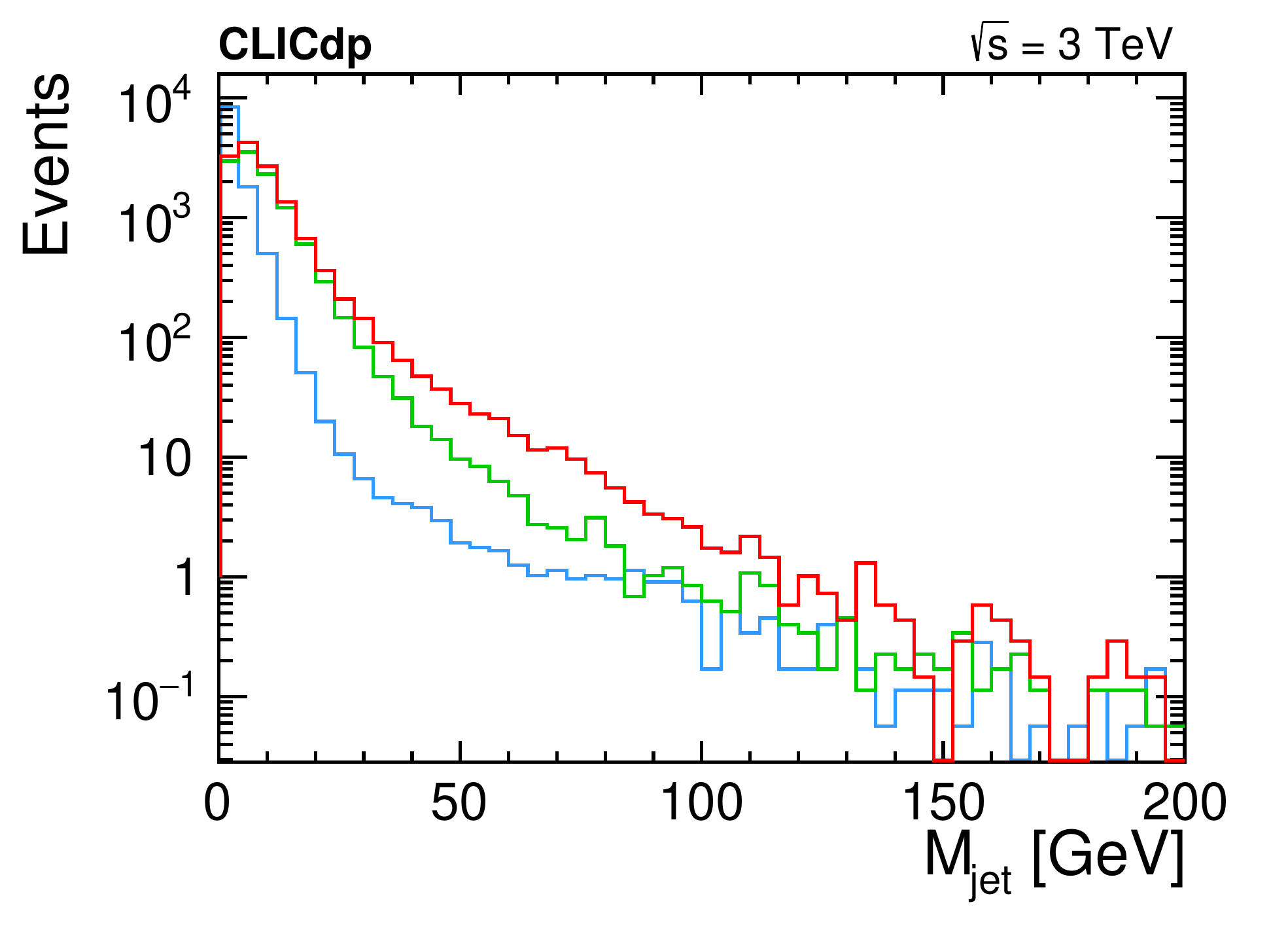}
	 	 	\label{ej1ej21500overlay}
	 	 \end{subfigure}
	 	 \begin{subfigure}{.5\textwidth}
	 	 	\centering
	 	 	\includegraphics[width=\linewidth]{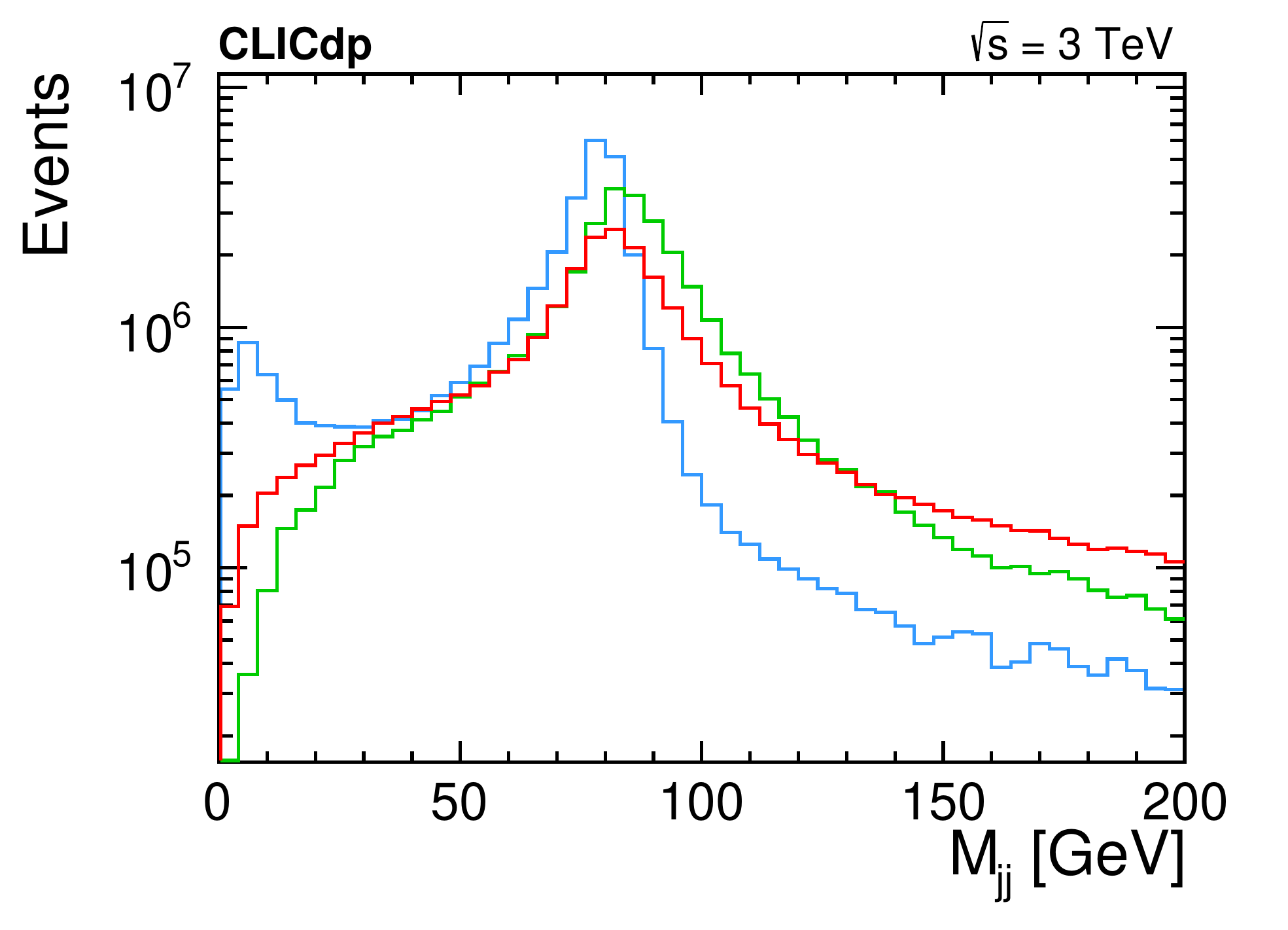}
	 	 	\label{mjj3000overlay}
	 	 \end{subfigure}%
	 	 \begin{subfigure}{.5\textwidth}
	 	 	\centering
	 	 	\includegraphics[width=\linewidth]{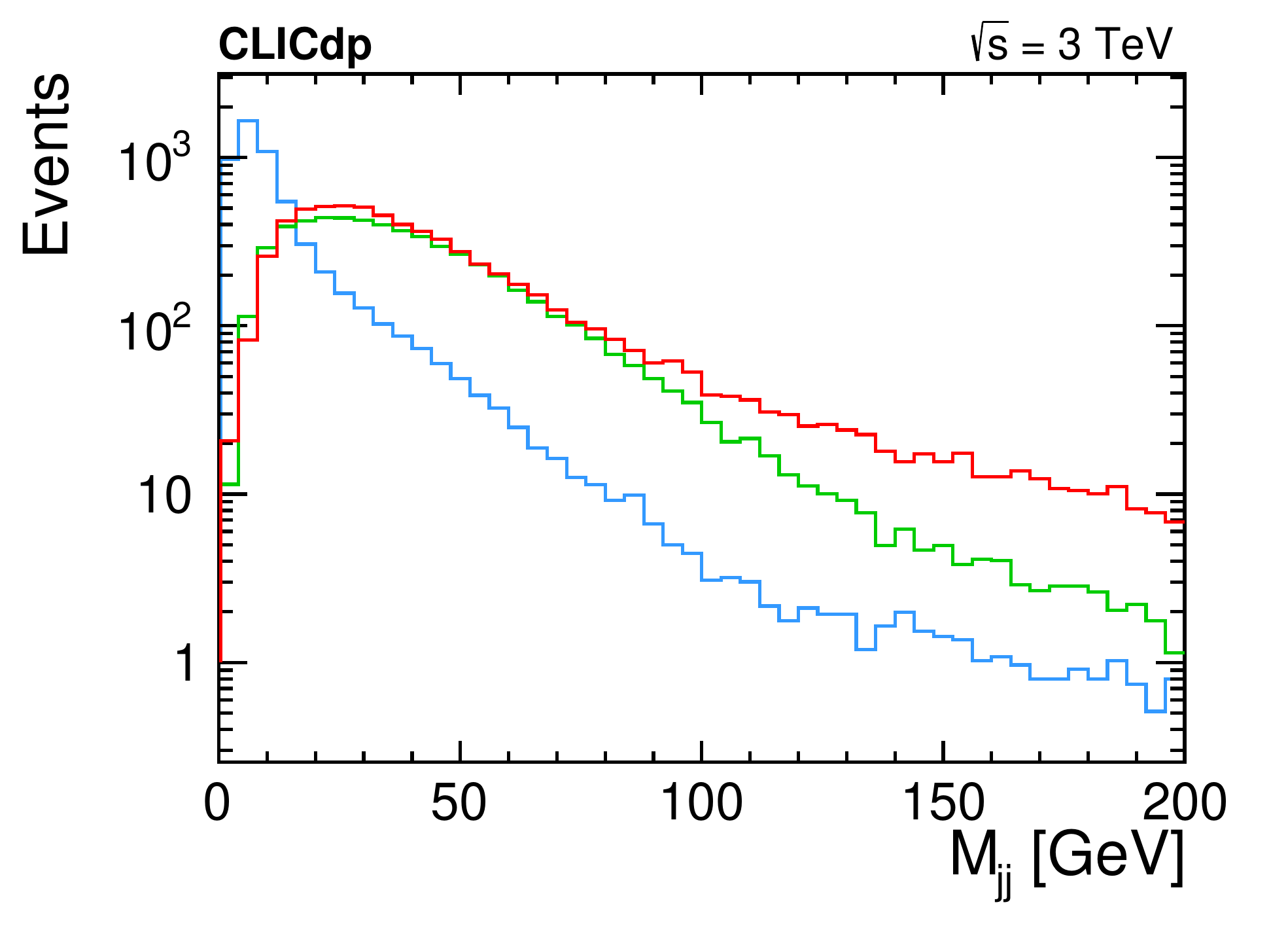}
	 	 	\label{ej1ej23000overlay}
	 	 \end{subfigure}
\caption{Histograms of the masses of a single jet (upper row) and of a dijet system (bottom row), for the qq$\ell\nu$ background (left) and HP17 signal (right) samples
  at \thirdstage. 
Results of the \delphes
  simulation without (blue histogram) and with (green) the overlay event contribution
  are compared with results of the full detector simulation (red). }
	 	 \label{fig:overlay}
	 \end{figure}

Figure \ref{fig:overlay} presents the impact of \ggtohad events produced in the
fast simulation on the distributions of variables describing jets. The
histograms resulting from \delphes with and without influence of the
overlay are compared to the outcome of the full simulation. 
Distributions are shown for the leading background channel, qq$\ell\nu$	(dominated by $W^+W^-$ production channel) and for the example signal scenario, HP17, with low scalar mass difference, $m_{H^\pm}-m_H \approx 10$\,GeV (see Tab.~\ref{tab:bmhigh}). $W$ mass peak is clearly visible in the di-jet mass distribution for the background, while the signal is dominated by low di-jet masses. Without the overlay event contribution, di-jet invariant mass distribution for signal is peaked, as expected, at around 10\,GeV. However, overlay background contribution strongly affects the measurement of the soft final state: the di-jet mass distribution gets significantly wider and is shifted towards higher values.
The improvement in the agreement between the fast and full simulation methods after
including the \ggtohad process in \delphes is clearly visible.
Good agreement is obtained for the maxima of the
 signal distributions, although contribution of events in high mass tails is 
  still underestimated.
  For background events, the maximum of the single jet invariant mass distribution is also well reproduced with fast simulation procedure, when overlay events are taken into account.
  Only the di-jet mass distribution for background events show systematic differences between fast and full simulation.
  We decided to apply the described selection, reflecting the timing cuts used in the full
  simulation, as only marginal improvement could be achieved by fine-tuning the
  procedure and cuts applied to the overlay events.

\section{Event preselection \label{sec:presel}}

As a first step,
only events with exactly one isolated lepton, electron or muon,
expected from a leptonic $W$ boson decay, and a pair of jets (di-jet
system) resulting from a hadronic decay of a second W boson were selected,
corresponding to the considered signal signature.  
Processes with tau lepton production were also included, both for
signal and background samples. However, most of these events were rejected at
this stage, as tau jet tagging was not used and only leptonic tau decays
could match the required event topology.
Furthermore, to avoid possible bias due to hard initial state
radiation, events containing at least one isolated photon with
energy $E_{\gamma}>10$\,GeV were rejected.
Also, the total transverse momentum of PFOs not contributing to the
required final state (two jets and a lepton), $p_{T}^{utg}$
(\textit{untagged} transverse momentum) had to be smaller than 20 GeV. 
This cut was imposed to reject events with significant deposits
  excluded from the reconstructed final state in the VLC clustering (activities
  in the forward direction assigned to the beam jets). 

As a next step, kinematic variables describing the event were
calculated and a simple cut-based preselection was applied.
The criteria used in the analysis for the two considered CLIC running
stages are presented in Table~\ref{tab:preselec_cuts}.
\begin{table}[tbp]
\centering
\hspace*{3 cm}
\caption{Preselection cuts applied to the kinematic
  variables calculated for selected signal and background events
         at $\sqrt{s} = 1.5$\,TeV and $\sqrt{s} = 3$\,TeV. } 
\begin{tabular}{|c|c|c|c|} \hline

variable & cut @ 1.5\,TeV & cut @ 3\,TeV \\ \hline
$E_{jj}$ & $< 750\,\mathrm{GeV}$ & $< 1.3\,\mathrm{TeV}$  \\
$M_{jj}$ & $> 3\,\mathrm{GeV}$ & $> 3\,\mathrm{GeV}$  \\
$\theta_{jj}$ & $0.2 < \theta_{jj} < \pi - 0.2$ & $0.3 < \theta_{jj} < \pi - 0.3$  \\
$E_{\ell}$ & $< 600\,\mathrm{GeV}$ & $< 1\,\mathrm{TeV}$  \\
$p_{T}^{\ell}$ & $< 550\,\mathrm{GeV}$ & $< 800\,\mathrm{GeV}$  \\
$\theta_{\ell}$ & $0.25 < \theta_{\ell} < \pi - 0.25$ & $0.5 < \theta_{\ell} < \pi - 0.5$  \\
$M_{miss}$ & $> 400\,\mathrm{GeV}$ & -  \\ \hline

\end{tabular}

\label{tab:preselec_cuts}
\end{table}
The following variables were used in the procedure:
\begin{itemize}\setlength\itemsep{-0.2em}  
\item $E_{jj}$ - energy of a dijet system,
\item $M_{jj}$ - invariant mass of a dijet system\footnote{The cut on a dijet system mass was introduced to suppress significant contributions of leptonic events misidentified as two-jet events in the fast simulation analysis, when the overlay background was not taken into account. This cut is not required when the overlay background contribution is included nor in the full simulation analysis. It was kept for the consistency.}, 
\item $\theta_{jj}$ - polar angle of a dijet system,
\item $E_{\ell}$ - energy of an isolated lepton,
\item $p_{T}^{\ell}$ - transverse momentum of an isolated lepton,
\item $\theta_{\ell}$ - polar angle of an isolated lepton,
\item $M_{miss}$ - missing mass, which is an invariant mass of
the missing four-momentum, $P_{miss}$, calculated by subtracting four-momenta of
a lepton and two jets from the four-momentum ($\vec{p}=\vec{0}$, $E =
\sqrt{s}$) of the initial state. 
\end{itemize}

%
%

\begin{figure}[tb]
	    \centering
	 	 \begin{subfigure}{.5\textwidth}
	 	 	\centering
	 	 	\includegraphics[width=\linewidth]{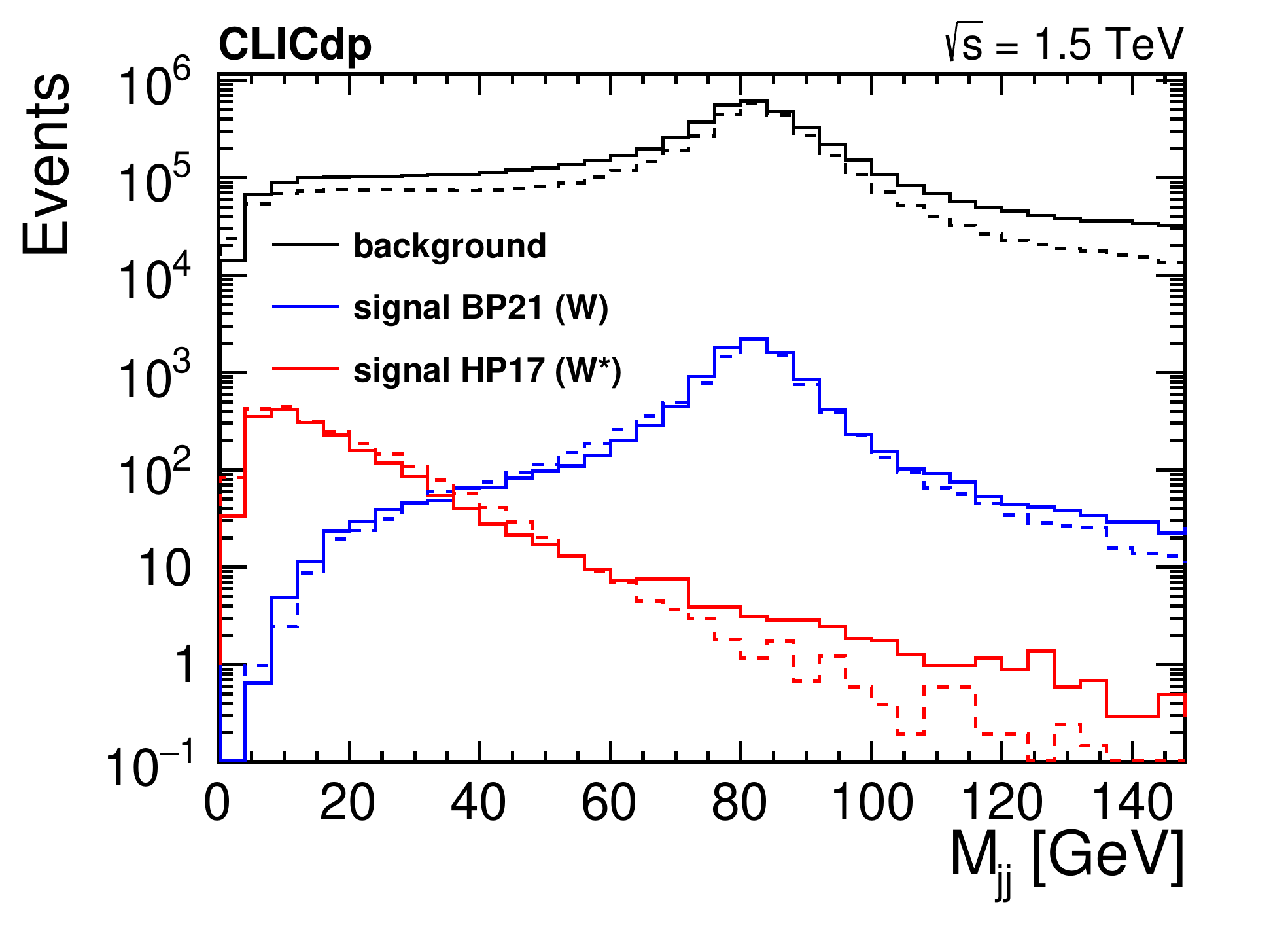}
	 	 	\label{mjj1500full_simulation}
	 	 \end{subfigure}%
	 	 \begin{subfigure}{.5\textwidth}
	 	 	\centering
	 	 	\includegraphics[width=\linewidth]{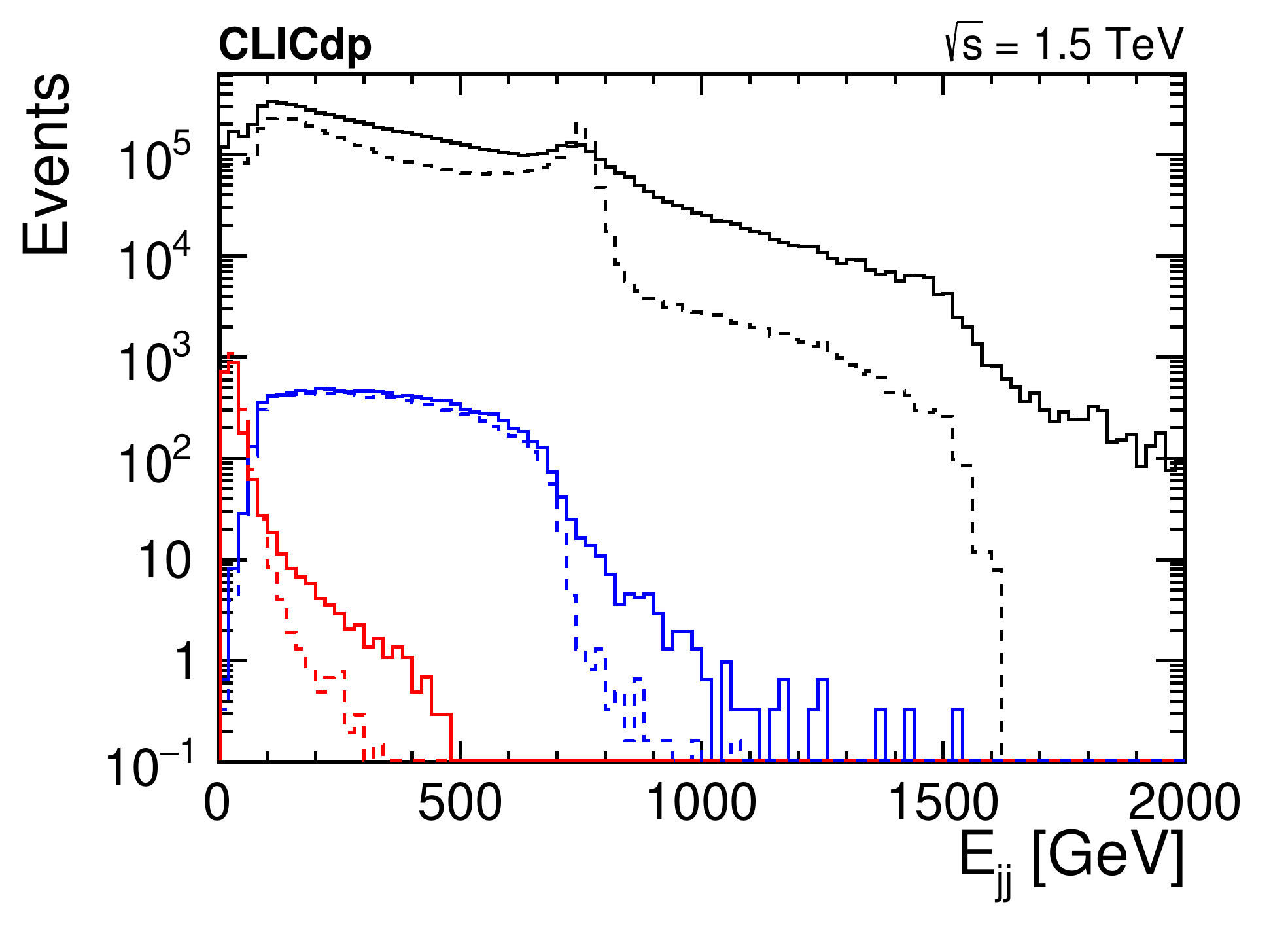}
	 	 	\label{ej1ej21500full_simulation}
	 	 \end{subfigure}
	 	 \begin{subfigure}{.5\textwidth}
	 	 	\centering
	 	 	\includegraphics[width=\linewidth]{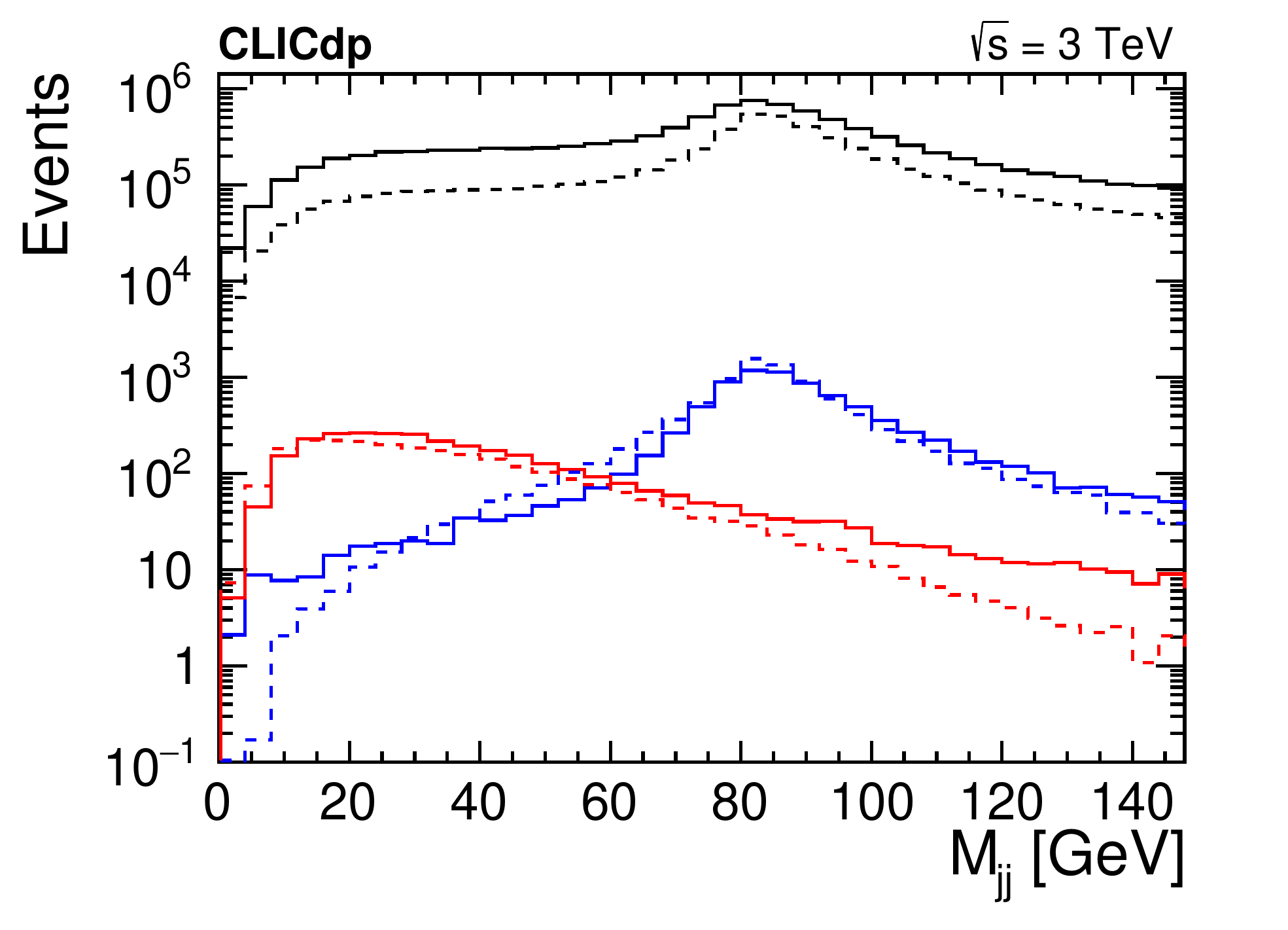}
	 	 	\label{mjj3000full_simulation}
	 	 \end{subfigure}%
	 	 \begin{subfigure}{.5\textwidth}
	 	 	\centering
	 	 	\includegraphics[width=\linewidth]{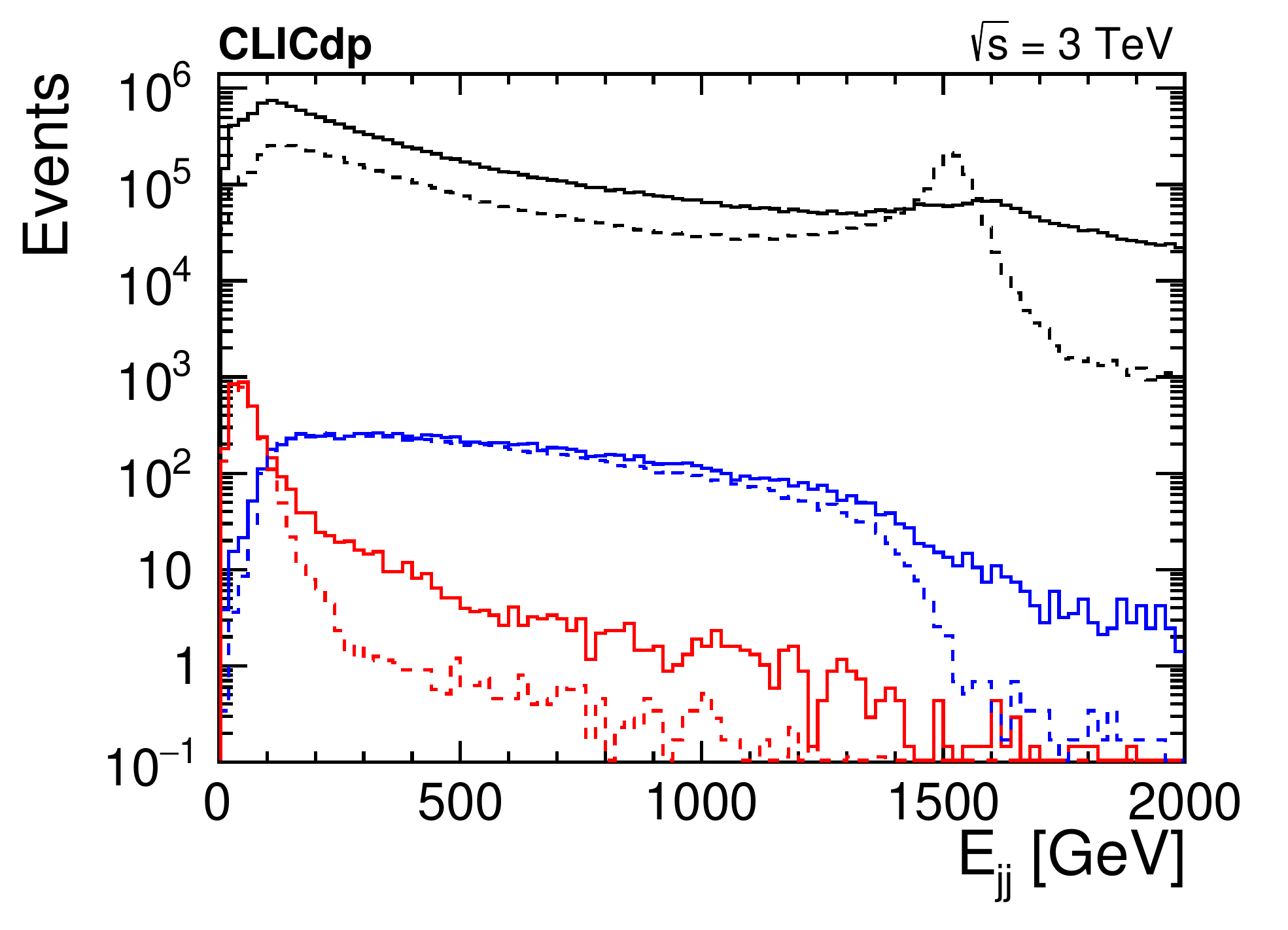}
	 	 	\label{ej1ej23000full_simulation}
	 	 \end{subfigure}
\caption{Histograms of the masses (left) and the energies (right) of a dijet system 
  at \secondstage (upper) and \thirdstage (bottom). The red line denotes HP17, and the blue one
  BP21. The black histogram is the sum of all background
  channels. Histograms are normalized to the number of events expected
  in the real experiment. Solid line shows the full simulation
  histograms and the dashed line, the corresponding fast simulation
  ones.} 
	 	 \label{fig:distributions}
	 \end{figure}

Figure \ref{fig:distributions}
shows the distributions of two of the variables used for the preselection cuts
at \secondstage and \thirdstage CLIC.
%
%
Presented are histograms for the
combined SM backgrounds and for the two example signal scenarios: one
with on-shell W$^\pm$ boson production (BP21) and one where the produced
W$^\pm$ is far off-shell (HP17).
Distributions obtained with full and fast detector simulation and
reconstruction are compared.
Despite some discrepancies, general agreement in the shapes of the
distributions is observed, which is not the case for \delphes without
including the \ggtohad overlay background.
In particular, good agreement is obtained in the dijet mass
distributions, which is very sensitive to the overlay contribution.
All histograms are normalised to the numbers of events expected at
CLIC, for nominal integrated luminosity.
Differences in normalisation between the fast and full simulation reflect the
differences in preselection efficiencies.
Higher background level estimates resulting from full simulation are
mainly due to the contribution from false lepton identification (events
without a final state lepton on the parton level), which is not modeled in
\delphes. 

	 \begin{table}[p]
			\centering
			\hspace*{3 cm}
\caption{Results of the event preselection at $\sqrt{s}=1.5$\,TeV CLIC
  running stage for the benchmark points and background channels
  considered in the  full simulation study. Shown are 
  cross sections $\sigma$, numbers of events expected after
  preselection cuts for an integrated luminosity of 2\,ab$^{-1}$
  and the preselection efficiency. For signal scenarios, the mass scale of the
  charged IDM scalar is indicated in parenthesis, for other model parameters
  see Table~\ref{tab:bmhigh} in the Appendix. Signal scenarios with off-shell 
  W$^\pm$ production are marked with a star.
} 
			\centering
			\small
            \begin{tabular}{|c|c|c|c|c|} \hline
channel ($m_{H^{\pm}}$ [GeV]) & $\sigma$ [fb] & exp. ev. after pres. & eff.  \\ \hline
BP21 \; (300)	&	8.09	&	9110	&	56\% \\
BP23 \; (191)	&	12.5	&	14200	&	57\% \\
HP17$^{*}$ (633)	&	2.43	&	1600	&	33\% \\
HP20$^{*}$ (441)	&	1.32	&	1500	&	57\% \\
HP3$^{*}$ \; (544)  &	0.629	&	749	&	60\%         \\	 \hline
        	qq$\ell\nu$	&	7000	&	1020000	&	7.3\% 	\\	
        	qq$\ell\ell$	&	2715	&	244000	&	4.5\% 	\\		
        	$\ell\ell$	&	1406	&	140000	&	5\% 	\\		
        	qqqq	&	1940	&	14100	&	0.36\% 	\\		
        	qq$\ell\nu\ell\nu$	&	14.9	&	2370	&	8\% 	\\		
        	qqqq$\ell\nu$	&	169	&	14400	&	4.3\% 	\\		
        	qq$\ell\nu\nu\nu$	&	66.8	&	56100	&	42\% 	\\		
        	qq$\nu\nu$	&	1460	&	124000	&	4.2\% 	\\		
        	qq	&	2860	&	48000	&	0.84\% 	\\		\cline{1-4}
            tot.	backg.	&	17638.9	&	1660000	&	4.7\% 	\\		\cline{1-4}
            \end{tabular}

			\label{tab:preselec_fullsim_1500}
	\end{table}
	 \begin{table}[p]
			\centering
			\hspace*{3 cm}
\caption{Results of the preselection at $\sqrt{s}=3$\,TeV CLIC running
  stage for the benchmark points and background channels considered
  in the  full simulation study. Shown are cross sections $\sigma$,
  numbers of events expected after preselection cuts for an integrated
  luminosity of 4\,ab$^{-1}$  and the preselection efficiency. 
  For signal scenarios, the mass scale of the
  charged IDM scalar is indicated in parenthesis, for other model parameters
  see Table~\ref{tab:bmhigh} in the Appendix. Scenarios with off-shell
  W$^\pm$ production are marked with a star.} 
			\centering
			\small
            \begin{tabular}{|c|c|c|c|} \hline
            channel ($m_{H^{\pm}}$ [GeV]) & $\sigma$ [fb] & exp. ev. after pres. & eff. \\ \hline
			BP21 \;	(300)	&	4.21	&	6460	&	38\%  \\		
            BP23 \; (191)	&	5.77	&	8490	&	37\% 	\\
            HP17$^{*}$	(633)	&	1.68	&	2020	&   30\%	\\	
            HP20$^{*}$	(441)	&	1.51	&	2640	&	44\% 	\\	
            HP3$^{*}$ \; (544)	&	1.78	&	3150	&	44\% 	\\ \hline
        	qq$\ell\nu$	&	8670	&	707000	&	2\% 	\\	
        	qq$\ell\ell$	&	3180	&	203000	&	1.6\% 	\\	
        	$\ell\ell$	&	1670	&	83000	&	1.2\% 	\\	
        	qqqq	&	902	&	14900	&	0.41\% 	\\	
        	qq$\ell\nu\ell\nu$	&	20.4	&	6610	&	8.1\% 	\\	
        	qqqq$\ell\nu$	&	148	&	17300	&	2.9\% 	\\	
        	qq$\ell\nu\nu\nu$	&	96.8	&	72500	&	19\% 	\\
        	qq$\nu\nu$	&	2330	&	212000	&	2.3\% 	\\	
        	qq	&	1270	&	36400	&	0.72\% 	\\ \cline{1-4}
            tot.	backg.	&	18286.7	&	1350000	&	1.8\% 	\\	\cline{1-4}
            \end{tabular}

			\label{tab:preselec_fullsim_3000}
	\end{table}

Results of the preselection
for all background channels
and signal scenarios considered for the full simulation are presented
in Tables~\ref{tab:preselec_fullsim_1500} and
\ref{tab:preselec_fullsim_3000}, for \secondstage and \thirdstage CLIC
stages, respectively.
Presented are the generator level cross sections, $\sigma$, number of
events expected after preselection,
assuming $-80$\% electron beam polarisation and total
integrated luminosities of 2\,ab$^{-1}$ (\secondstage) and
4\,ab$^{-1}$ (\thirdstage),
and the corresponding preselection efficiencies.
The largest contribution to the SM background after preselection comes
from $qql\nu$ production (dominated by pair production of $W$ bosons),
mainly due to the largest generator level cross section and the
matching event topology.
The highest preselection efficiency (weakest background suppression) is
obtained for the qq$\ell\nu\nu\nu$ final state, which matches the
expected signal topology as well and also results in large
reconstructed missing mass from the three escaping neutrinos.
Fortunately, this background channel has a relatively small cross section.

%

\section{Multivariate analysis \label{sec:mva}}

After the cut-based preselection, Boosted Decision Trees (BDTs) were
used for the multivariate analysis, as implemented in the \tmva
toolkit~\cite{hoecker2007tmva}. The classifier consisted of 1000
decision trees and a principal component analysis was applied on
the input data in the preprocessing phase.
Decision trees were randomised, which means that every decision tree
uses only a randomly chosen subset of input variables (six in case of
this study). The following variables were used as an input to
the BDT algorithm, in addition to the set used in preselection (refer to
Sec.~\ref{sec:presel}):  
\begin{itemize}\setlength\itemsep{-0.2em}  
    \item $E_{j_1}$, $E_{j_2}$ - energies of the two jets,
    \item $p_{T}^{j_1}$, $p_{T}^{j_2}$ - transverse momenta of the two jets,
    \item  $\theta_{j_1}$, $\theta_{j_2}$- polar angles of the two jets,
    \item $\Delta\theta_{Wj}^{*}$ - polar angle between the leading (higher $p_T$) jet
      and the dijet direction (flight direction of
      hadronically decaying $W^{\pm}$), calculated in the dijet centre-of-mass frame,
    \item $\Delta\phi_{Wj}^{*}$ - azimuthal angle between the jet and
      the dijet direction, calculated in the dijet centre-of-mass frame,
    \item $p_{T}^{utg}$ - untagged transverse momentum,
    \item $E_{T}^{miss}$ - missing transverse energy (MET), calculated as the
transverse component of $P_{miss}$.
\end{itemize}

In the full simulation study, BDTs were trained separately for each of the
considered signal scenarios.
A more conservative approach would be to use many different signal
scenarios as an input to the BDT training, as it is very unlikely that one
of the analysed benchmark points is realised by nature.
Unfortunately,  the data-set available for the full simulation analysis was
limited to only five benchmark scenarios, two with on-shell and three
with off-shell W$^\pm$ production (these two classes of scenarios should
be trained separately).
A single scenario optimisation approach was chosen to avoid possible bias of
sensitivity towards one of the signal scenarios (or to particular region
in the parameter space).
The same approach was used when comparing full simulation results for
these 5 scenarios with results based on the \delphes fast simulation.

\begin{figure}[tb]
	    \centering
	 	 \begin{subfigure}{.45\textwidth}
	 	 	\centering
	 	 	\includegraphics[width=\linewidth]{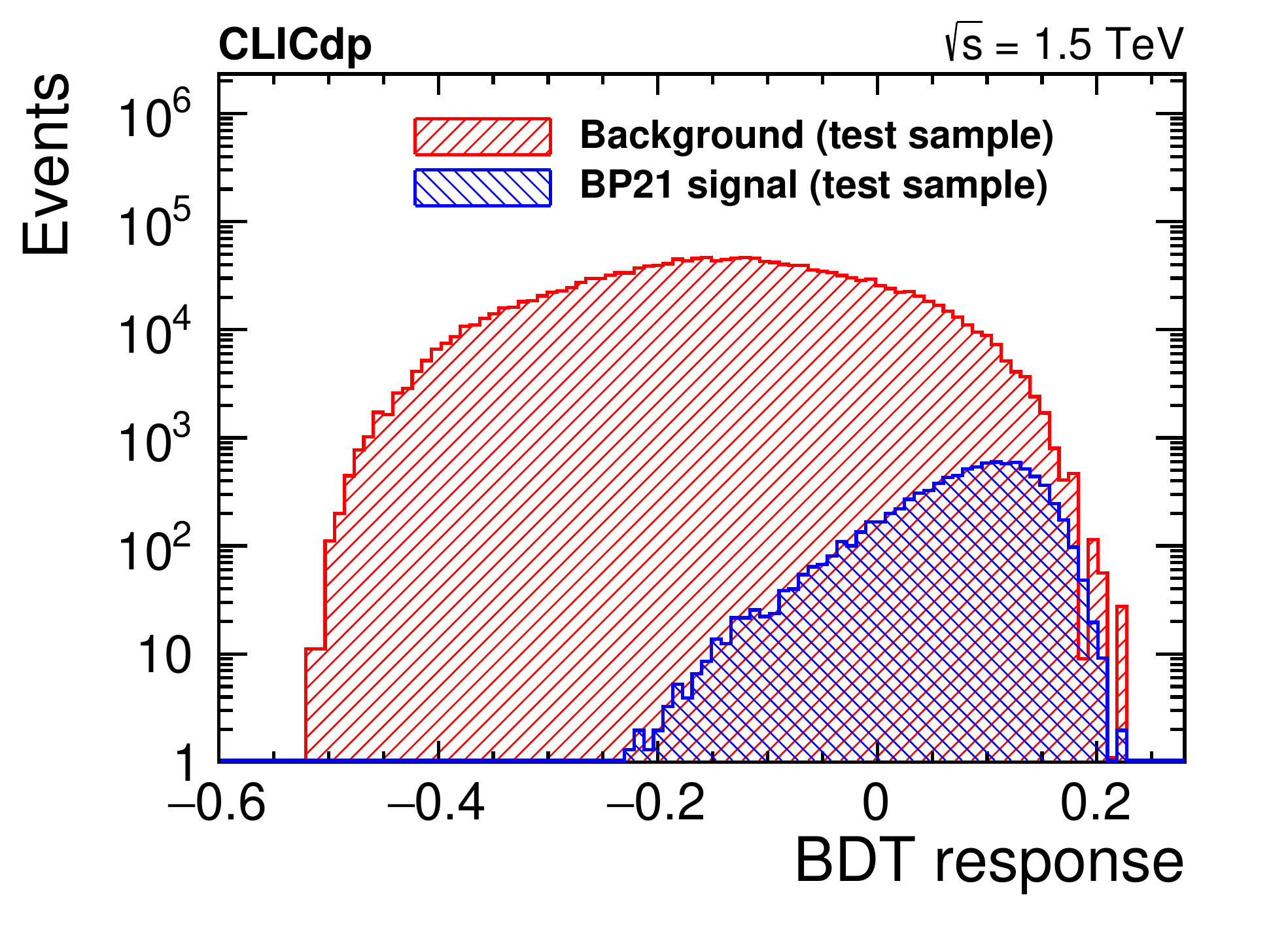}
	 	 \end{subfigure}%
	 	 \begin{subfigure}{.45\textwidth}
	 	 	\centering
	 	 	\includegraphics[width=\linewidth]{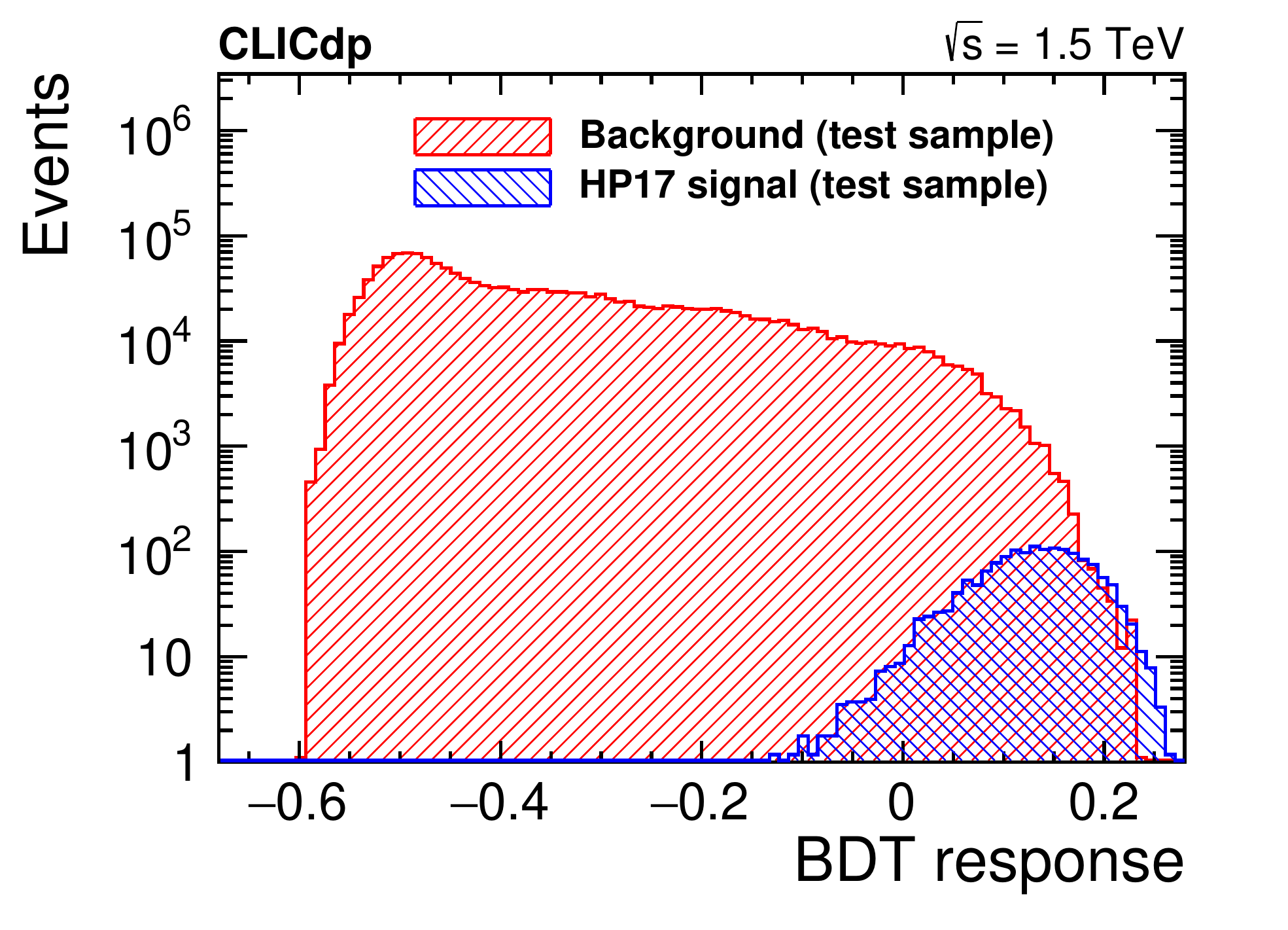}
	 	 \end{subfigure}
	 	 \begin{subfigure}{.45\textwidth}
	 	 	\centering
	 	 	\includegraphics[width=\linewidth]{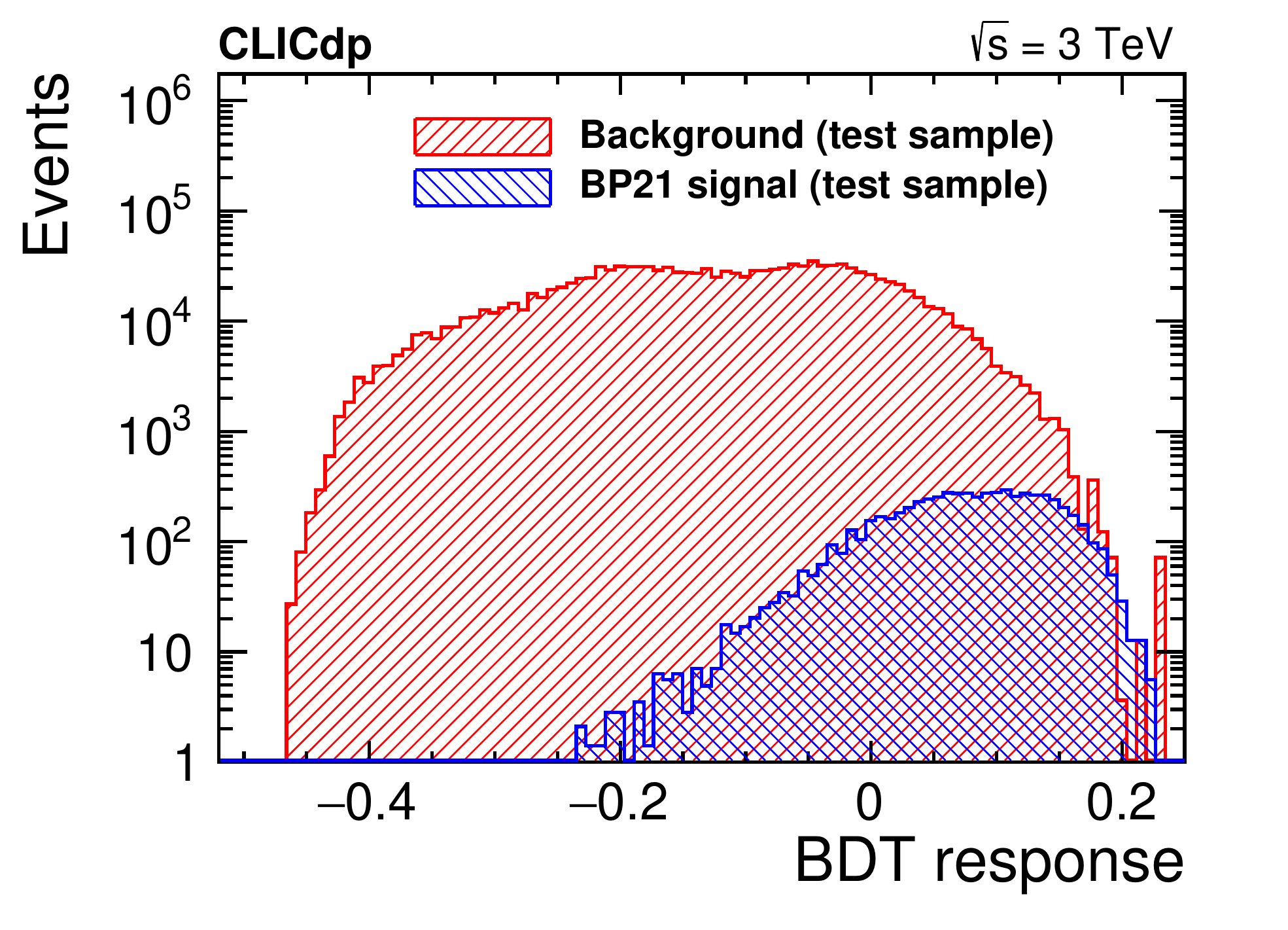}
	 	 \end{subfigure}%
	 	 \begin{subfigure}{.45\textwidth}
	 	 	\centering
	 	 	\includegraphics[width=\linewidth]{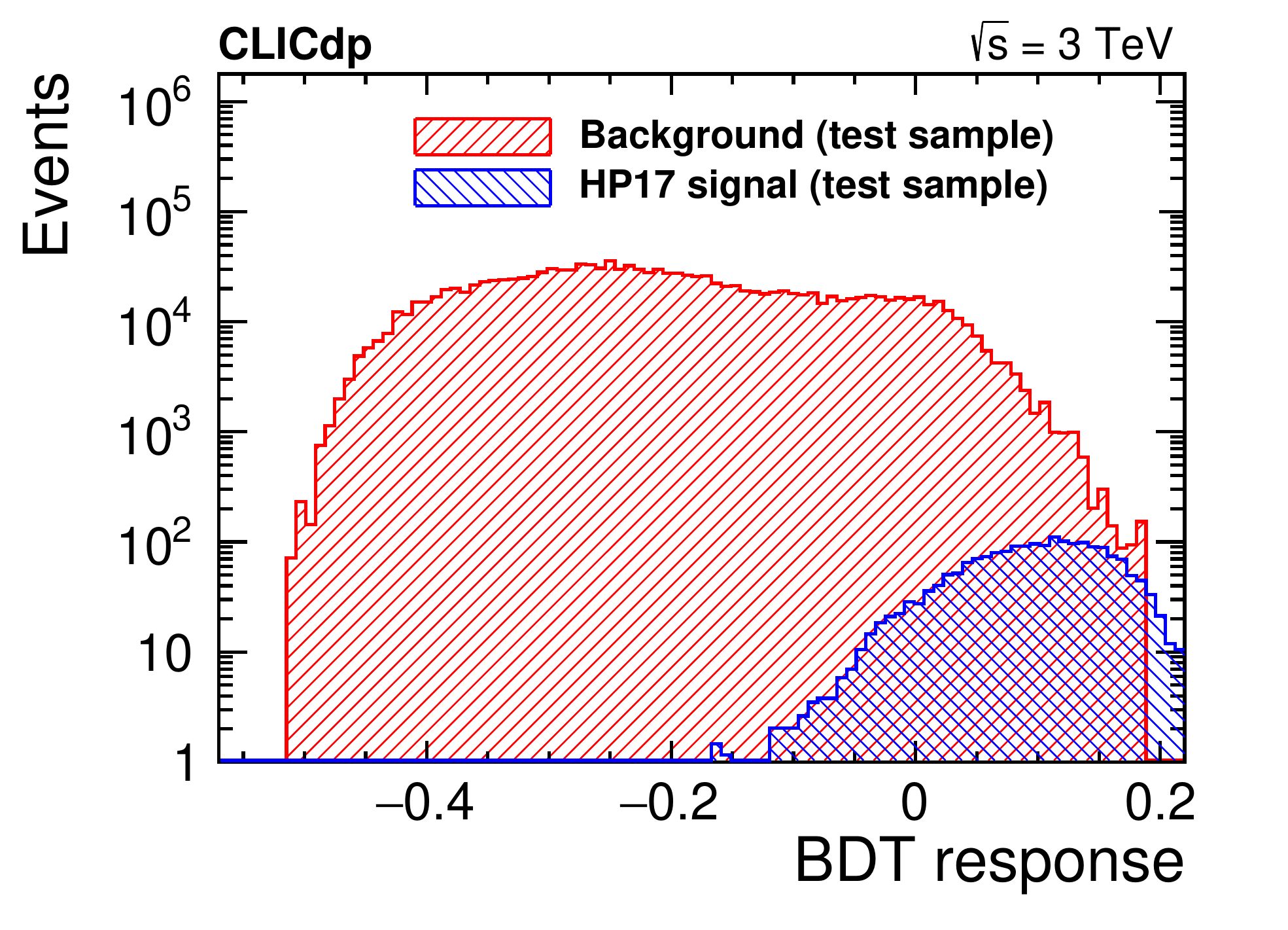}
	 	 \end{subfigure}
\caption{BDT response distributions for the BP21 (left) and the HP17
  (right) scenarios at $\sqrt{s} = 1.5$\,TeV (upper plots) and at
  $\sqrt{s} = 3$\,TeV (bottom plots). Benchmark scenarios and SM
  background expectations are normalised to the integrated luminosity
  of 2\,ab$^{-1}$ and 4\,ab$^{-1}$, for 1.5\,TeV and 3\,TeV,
  respectively. All of the presented distributions were obtained with
  full detector simulation and maximal achievable significance with
  them exceeds 5$\sigma$. } 
	 	 \label{bdtoutputFullSim}
\end{figure}

Example distributions of the BDT response are presented in
Figure~\ref{bdtoutputFullSim}, for the two selected signal scenarios,
BP21 and HP17, and the two CLIC running stages.
The final event selection is optimised  separately for each scenario by
finding the cut on the BDT response that maximises the statistical
significance of expected deviations from the SM predictions,
calculated as $S/\sqrt{S+B}$, where $S$ and $B$ are the numbers of
signal and background events after the cut, respectively. 
Although it is not possible to select a signal dominated sample even with a tight BDT response cut, significant deviations from SM predictions can be observed in the measured event distributions.
For example, signal for BP21 scenario can be observed at 3\,TeV CLIC with statistical significance of about 18$\sigma$: about 2640 signal events, 41\% of preselected sample, pass the BDT selection for optimal response cut of about 0.10, while background contribution is suppressed by a factor of about 70, to 19\,000 event. Resulting signal to background ratio is $S/B \approx 0.14$: 14\% excess in the number of selected events, relative to the SM predictions, is expected in the considered IDM scenario.

When estimating the CLIC sensitivity to the charged IDM scalar production for
all of the 23 benchmark scenarios considered, based on the fast
detector response simulation with \delphes, a conservative approach was
used and the BDTs were trained on all available samples, separately for
scenarios with on- and off-shell W$^\pm$ boson production.
However, the optimal cut on the BDT response was found separately for
each signal scenario.
This is justified by the fact that, in the actual experiment, the
measured and predicted BDT response distributions should be compared
and not just the number of events after the cut. 
For both fast and full simulation analyses, an additional requirement
was imposed when looking for the optimal BDT response cut.
Only cut values resulting in a total signal selection efficiency of
at least 10\% were allowed, to reduce the impact of statistical
fluctuations in the generated MC samples.

There are visible correlations between input variables, 
resulting from their definitions.
The highest correlations occur between $E_{jj}$ and
$E_{j_{i}}$, $\theta_{W}$ and $\theta_{j_{i}}$ ($i=1,2$), as well as $E_{\ell}$ and
$p_{T}^{\ell}$, reaching from 93\% to  97\%.
In general, correlations depend on the considered sample (signal scenario or background).
Only for few variable pairs
high correlations are observed for all signal scenarios and SM background.
In most cases, two variables highly correlated in one
training data-set (e.g. the signal sample in case of the full simulation analysis) are not necessarily that much correlated in another
one.
Differences in variable correlations were also observed between
fast and full simulation studies.
Because of the randomisation procedure mentioned above, correlations
between variables do not affect the efficiency of the BDT training.
By keeping all considered variables as an input to the BDT, a consistent
approach can be used for all scenarios, for both fast and full simulation.
It was verified that removing any of these variables results
in worse discrimination-power between signal and background events,
and a reduced sensitivity to IDM scalar production for at least some of
the scenarios.

\section{Results \label{sec:results}}

\begin{figure}[tb]
	    \centering
	 	 \begin{subfigure}{0.5\textwidth}
	 	 	\centering
	 	 	\includegraphics[width=\linewidth]{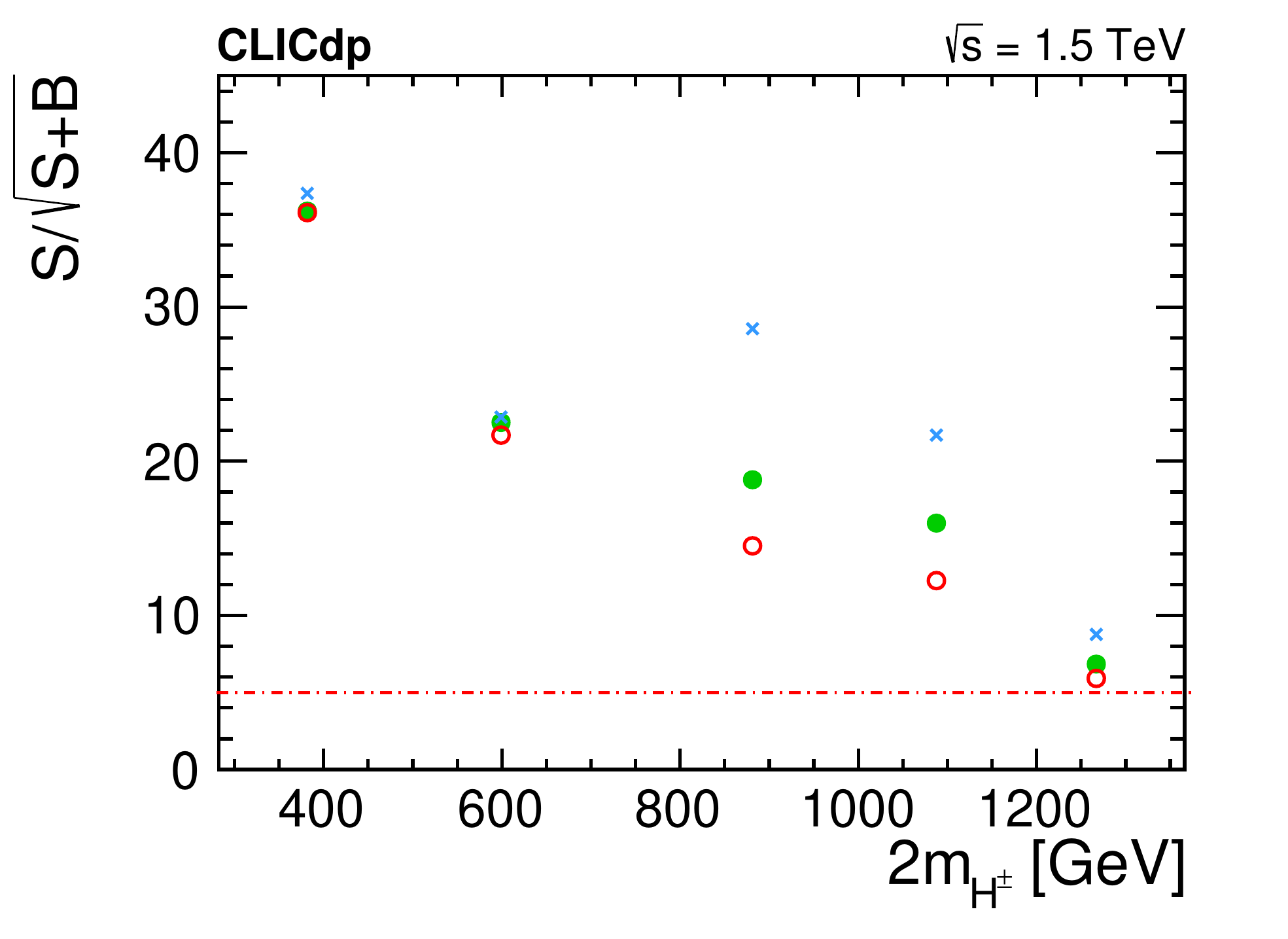}
	 	 \end{subfigure}%
	 	 \begin{subfigure}{0.5\textwidth}
	 	 	\centering
	 	 	\includegraphics[width=\linewidth]{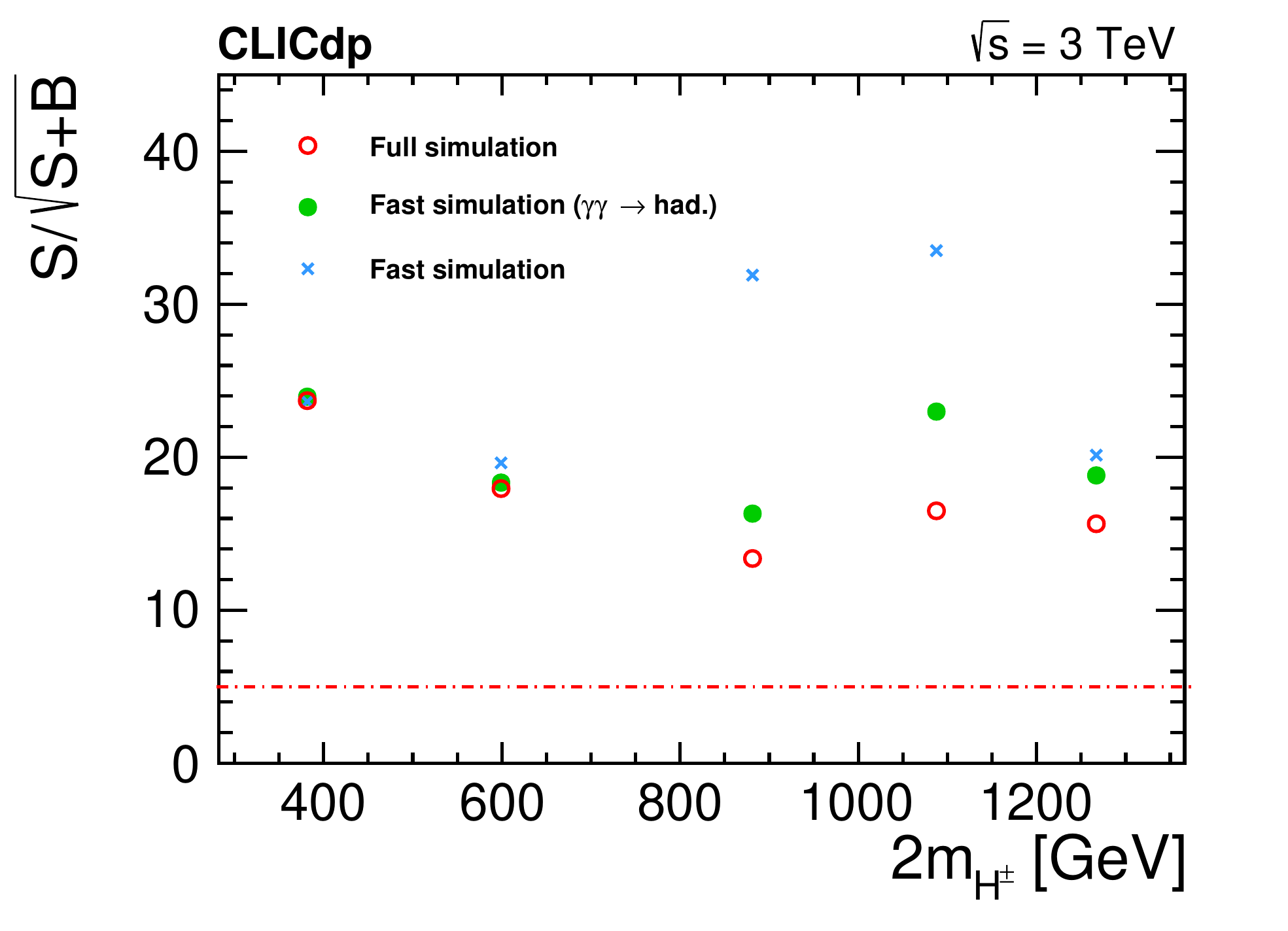}
	 	 \end{subfigure}
\caption{Statistical significance expected from the full simulation
  study (red points), compared with the fast simulation results after
  including the overlay background (green points). The significance
  obtainable from the fast simulation analysis without the influence of
  the overlay are shown for comparison as blue crosses. The results
  are presented for considered benchmark points as a function of
  $2m_{H^{\pm}}$, for 1.5\,TeV (left) and 3\,TeV (right) CLIC running,
  and were produced with the selection optimised for a particular
  scenario. The red line shows the 5$\sigma$ threshold.} 
	 	 \label{fig:resultsFastFullSim}
	 \end{figure}

The main goal of the analysis was to establish the statistical
significance of deviations from the SM predictions expected at high
energy CLIC stages for each of the considered benchmark scenarios.
The significance, calculated for the 5 scenarios included in the full
detector simulation analysis, is shown in
Figure~\ref{fig:resultsFastFullSim} as a function of the total mass of the
produced IDM scalars, which is twice the charged IDM scalar mass for
the considered process, $2 m_{H^\pm}$.
Results based on full simulation are compared with those based on
\delphes, with and without the \ggtohad background included.
All results in Figure~\ref{fig:resultsFastFullSim} were obtained with
the BDT selection optimised for an individual benchmark scenario.
While the expected significance is decreasing with  increasing scalar mass,
the pair production of charged IDM scalars can be discovered at CLIC
(significance above 5$\sigma$) for all benchmark points considered in the full simulation.
The results based on the fast simulation tend to be more optimistic.
However, when overlay background is included in the \delphes simulation,
the agreement with the full simulation results is significantly improved.

\begin{figure}[tb]
    	\centering
    	\includegraphics[angle=0,width=.65\textwidth]{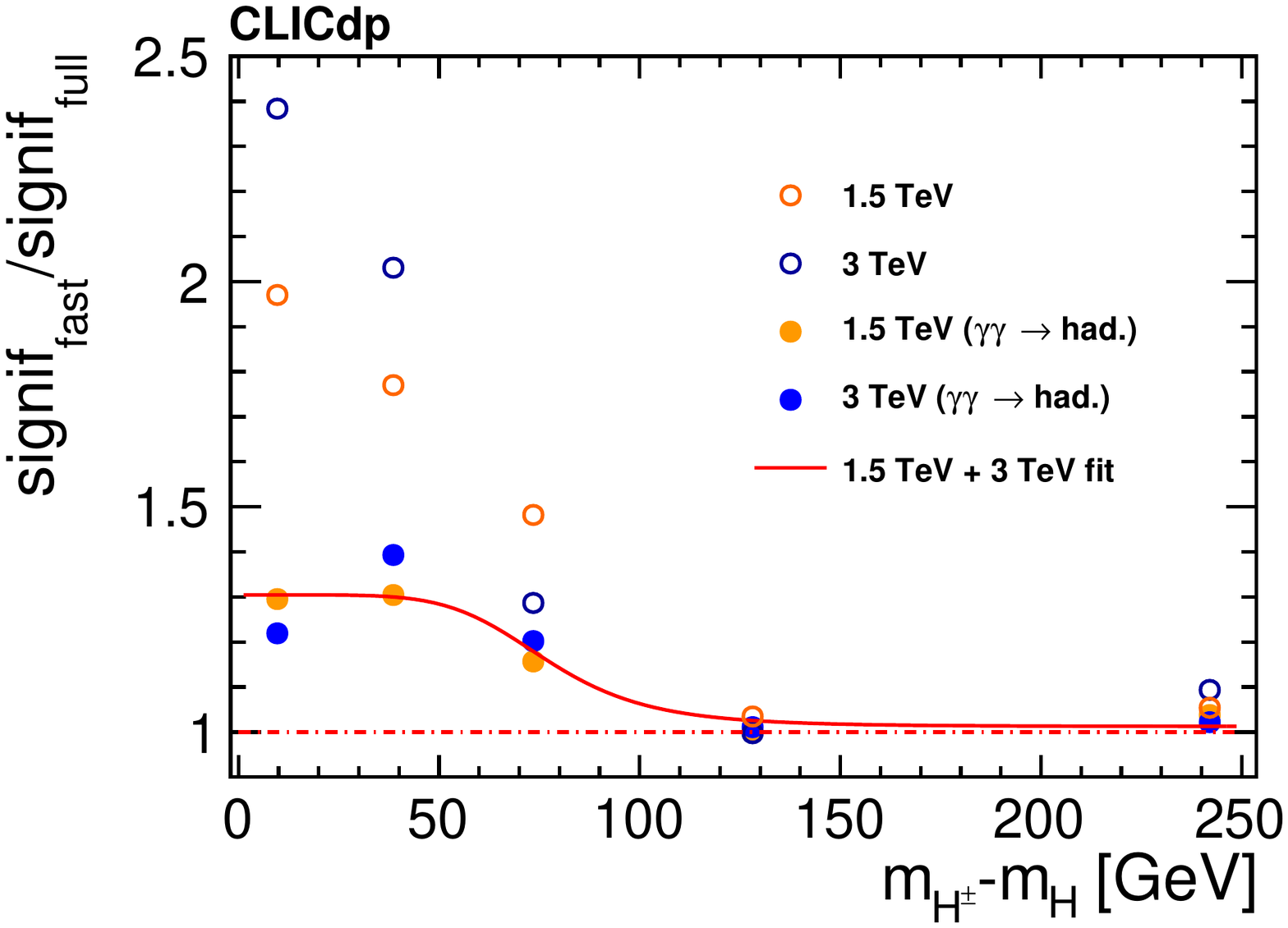}
\caption{Ratios of the expected observation significances resulting from
  fast and full simulation studies. Open circles correspond to \delphes\:simulation results with no
  overlaid \ggtohad events, while full circles show the ratios after
  inclusion of this background, for CLIC running at 1.5\,TeV (orange) and
  3\,TeV (blue points). The red curve corresponds to the dependence described by Eq.~(\ref{correction2}) used to correct for the remaining difference between full simulation results and \delphes with overlay at both energy stages. See the text for details.} 
    	\label{fig:resultsDiff}
    \end{figure}

The impact of the proposed procedure to include overlay contribution on
the agreement between fast and full simulation is also presented in
Figure~\ref{fig:resultsDiff}.
It shows the ratios of the significance obtained
using the fast simulation, with and without overlay background included,
to the full simulation (ratios of the significances presented in
Figure~\ref{fig:resultsFastFullSim}), as a function of the IDM scalar mass
difference, \chmassdiff, which determines the virtuality of the 
W bosons produced in the $H^\pm$ decays.
The two points with the highest mass splittings (BP21 and BP23)
correspond to the points with the smallest charged scalar mass $m_{H^\pm}$,
shown in Fig.~\ref{fig:resultsFastFullSim}.
For large mass differences, $m_{H^\pm} - m_H > m_W$, good agreement is
observed between fast and full simulation results.
If the overlay background is not included in \delphes, 
significant discrepancies between fast and full simulation arise
for low mass differences, $m_{H^\pm} - m_H < m_W$.
While the observed deviations are significantly reduced 
after the \ggtohad background is taken into account, some discrepancies
still remain.
Therefore, in order to further account for the possible systematic effects,
the correction factor $\Delta$ was introduced to describe the remaining discrepancy between significance, $\mathbf{S}$, resulting from fast and full simulation study:
\begin{equation}
 \label{correction}
   \mathbf{S}_\text{full} = \mathbf{S}_\text{fast} \cdot 
    \left[ 1 - \Delta(m_{H^{\pm}}-m_H) \right] \; ,
\end{equation}
where the correction $\Delta$, as suggested by Fig.~\ref{fig:resultsDiff}, is assumed to depend only on the scalar mass splitting. 
%
The following functional form was chosen to describe the correction:
\begin{equation}
\label{correction2}
    \Delta(\delta m) = a + \frac{b}{1+\left( \delta m / c \right)^6} \; ,
\end{equation}
where parameters  $a$, $b$ and $c$  
were fitted to the points corresponding to the fast simulation with
overlay background included, at both energy stages simultaneously.
The resulting value of parameter $c$ is consistent with the mass of the W boson, indicating that the differences between the full and fast simulation are relevant only when virtual boson production is expected, for \chmassdiff$ < m_W$.
When mass splitting is large and on-shell W production is expected for signal scenarios, similarly as for the dominant background channel (qq$\ell\nu$), possible differences between fast and full simulation do not affect signal-background separation and the results for the two simulation approaches are very similar.

\begin{figure}[tb]
	    \centering
	 	 \begin{subfigure}{.5\textwidth}
	 	 	\centering
	 	 	\includegraphics[width=\linewidth]{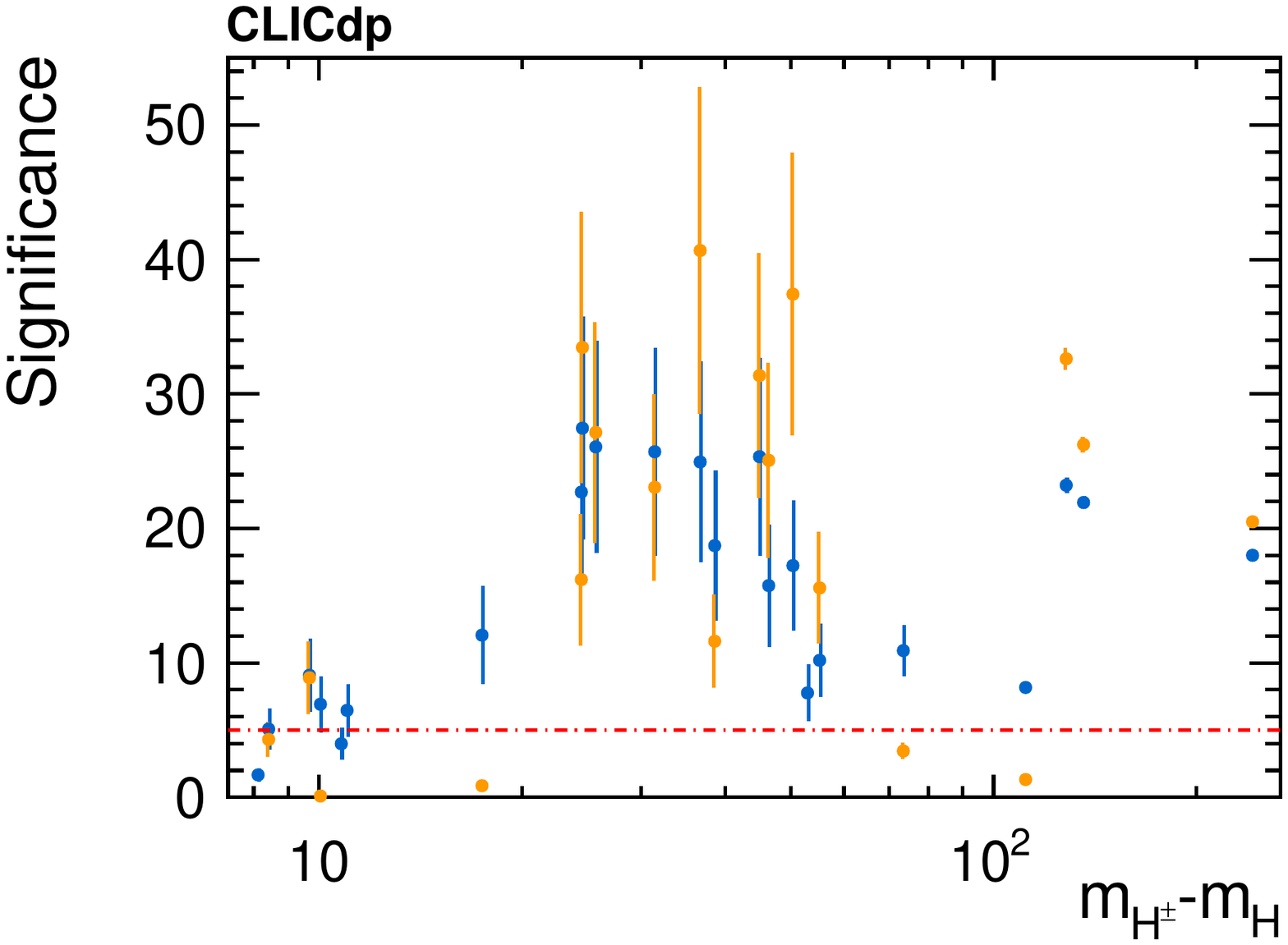}
	 	 \end{subfigure}%
	 	 \begin{subfigure}{.5\textwidth}
	 	 	\centering
	 	 	\includegraphics[width=\linewidth]{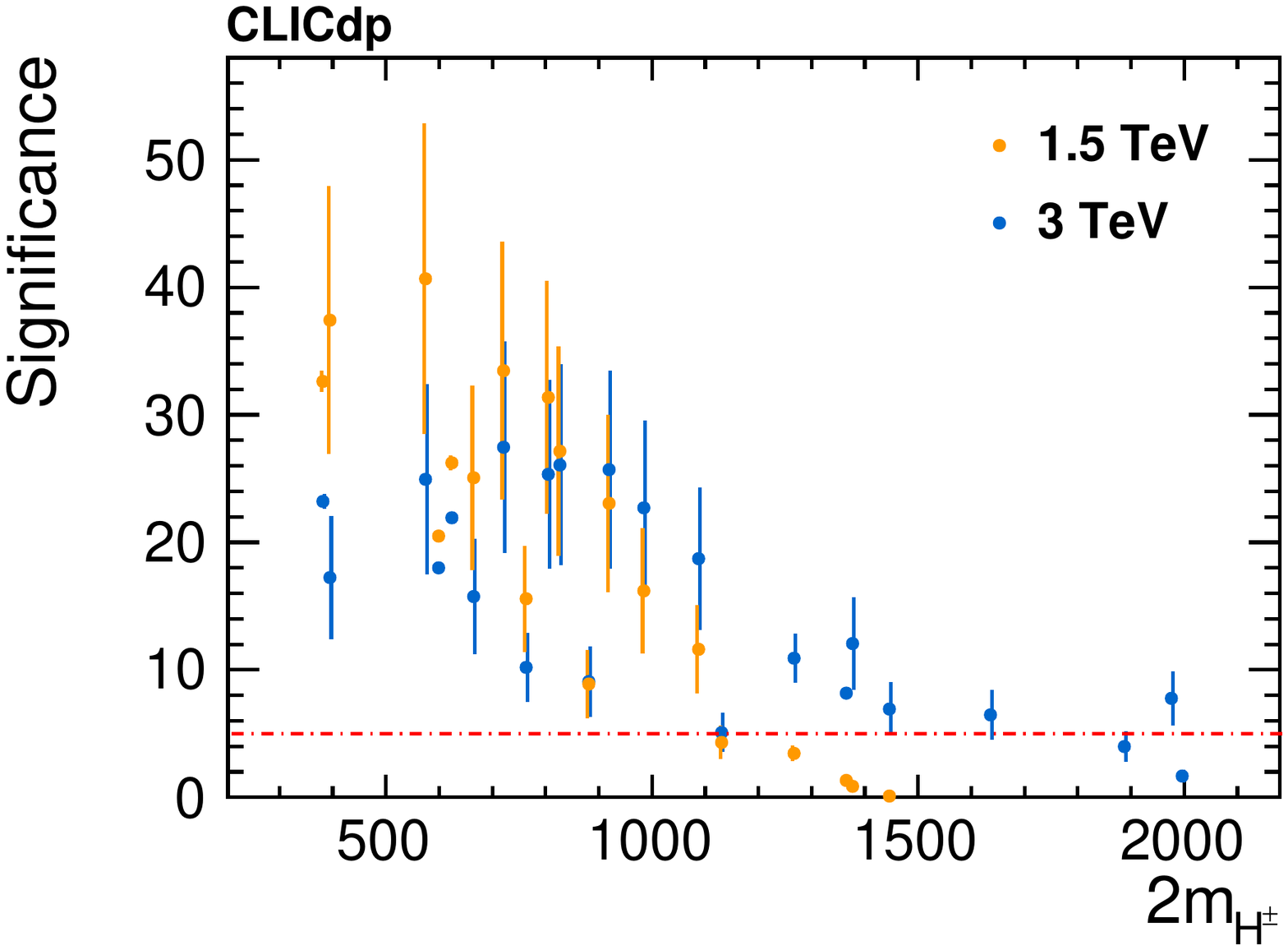}
	 	 \end{subfigure}
\caption{Expected statistical significance of IDM charged scalar
  pair-production observation as a function of the IDM scalar
  mass difference,  \chmassdiff (left) and of the
  total mass of the produced IDM scalars, $2m_{H^{\pm}}$ (right).
  Results of the \delphes fast simulation study are presented 
  for CLIC running at \secondstage (orange circles) and \thirdstage
  (blue points). Error bars indicate the systematic uncertainty, estimated from the observed difference between fast and full simulation results, see text for details. The red horizontal lines indicate the 
  5$\sigma$ threshold.} 
	 	 \label{fig:resultsFastSimAll}
	 \end{figure}

The correction given by Eq.~\ref{correction} was used to scale the significance results obtained with modified \delphes. To take other possible systematic effects and the arbitrary choice of the correction function into account, we assume 100\% uncertainty on $\Delta$.
Final results obtained for all considered benchmark scenarios with
the \delphes fast simulation, including overlay events and training the BDTs on all
available scenarios (as described in Sec.~\ref{sec:mva}), are presented in
Figure~\ref{fig:resultsFastSimAll} together with the uncertainties.
The expected statistical significance of IDM charged scalar production is
shown as a function of the scalar mass differences, \chmassdiff,
as well as of the total mass of the produced IDM scalars, $2\;
m_{H^\pm}$, for the two high energy stages of CLIC. 
For most of the considered scenarios, pair-production of charged
IDM scalars can be observed at CLIC with high significance, attaining
even more than 40$\sigma$, and for the scalar masses up to 1\,TeV.
Only for two of the benchmark points the expected significance is
below 5$\sigma$ and for three of them it is over 5$\sigma$ within less than
the systematic uncertainty.
There is also no visible dependence of the discovery reach on the dark
scalar mass difference but for scenarios with very small values of
\chmassdiff which are clearly much more challenging due to the
influence of \ggtohad events.

\begin{figure}[tb]
	    \centering
	 	 \begin{subfigure}{.5\textwidth}
	 	 	\centering
	 	 	\includegraphics[width=\linewidth]{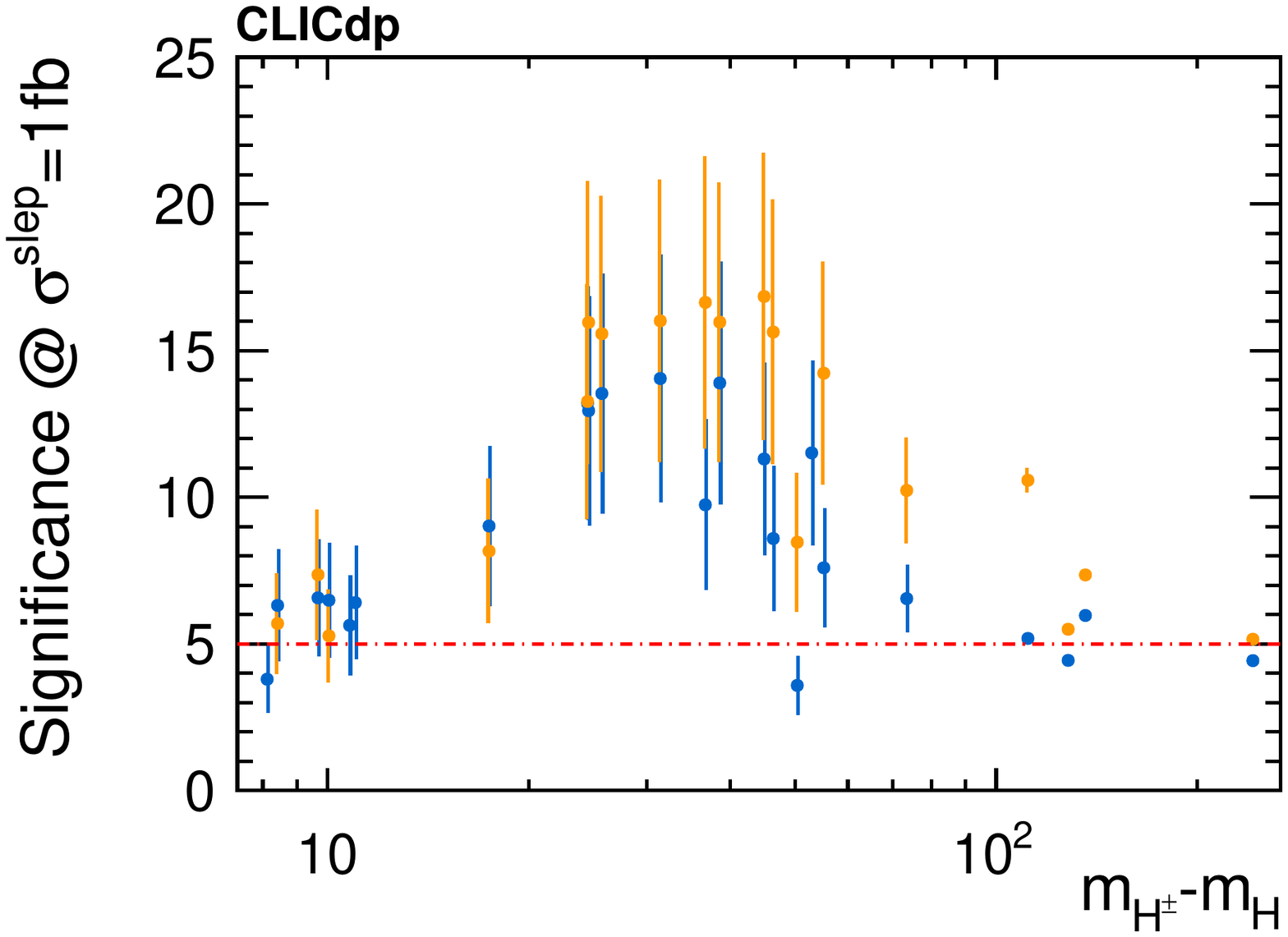}
	 	 \end{subfigure}%
	 	 \begin{subfigure}{.5\textwidth}
	 	 	\centering
	 	 	\includegraphics[width=\linewidth]{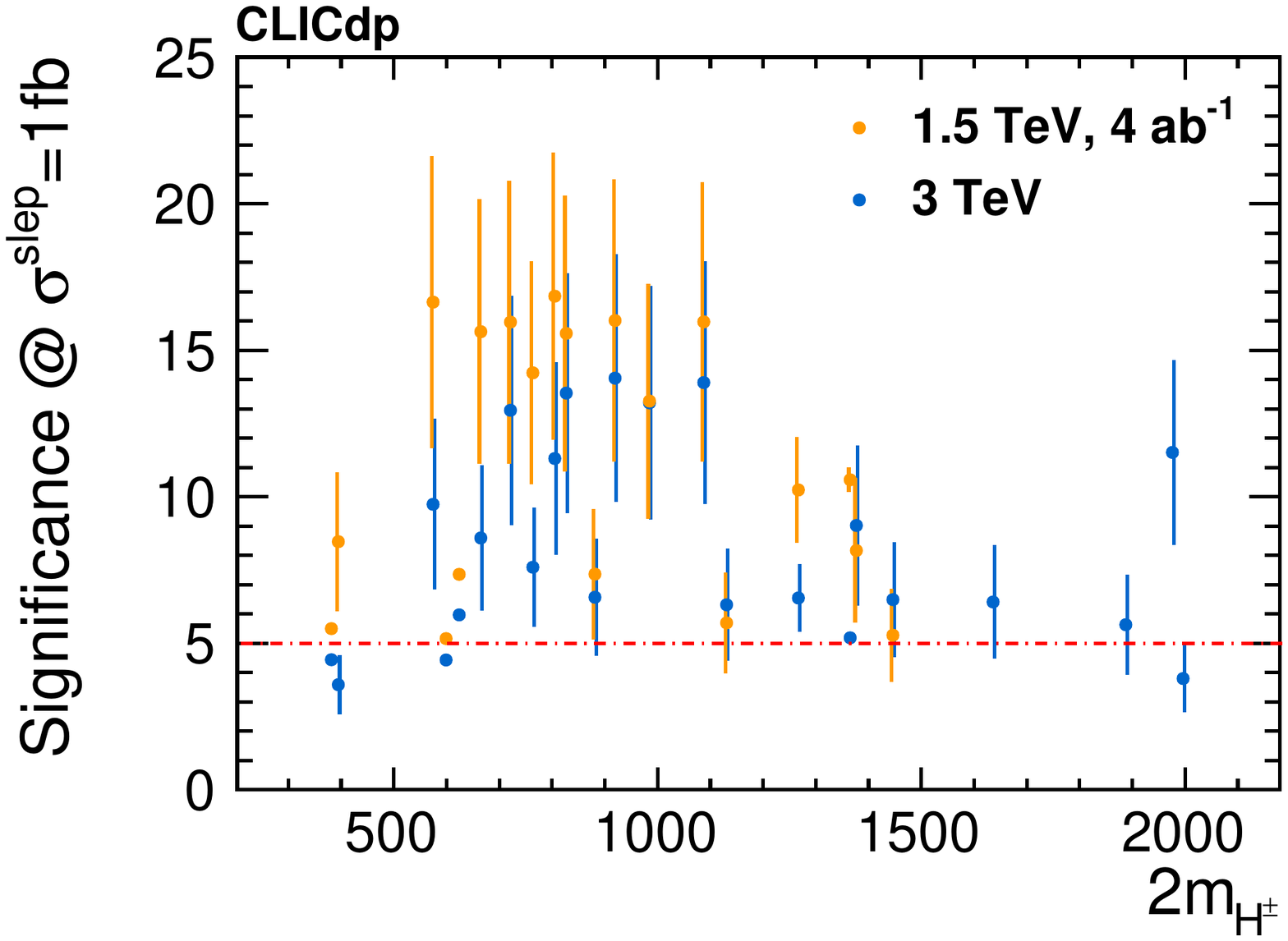}
	 	 \end{subfigure}
\caption{ Expected statistical significance of IDM charged scalar
    pair-production observation, assuming the semi-leptonic channel cross section of 1\,fb:  
  as a function of the IDM scalar mass difference,  \chmassdiff (left) 
  and of the total mass of the produced IDM scalars, $2m_{H^{\pm}}$ (right). 
  Results of the \delphes fast simulation study are presented 
  for CLIC running at \secondstage (orange circles) and \thirdstage
  (blue points) with 4\,ab$^{-1}$ of integrated luminosity assumed for both stages. Error bars indicate the systematic uncertainty estimated from the observed difference between fast and full simulation results, see text for details.
  The red horizontal lines indicate the  5$\sigma$ threshold. } 
	 	 \label{fig:resultsFastSimScaled}
	 \end{figure}

The significance of the observation is expected to depend
  mainly on the signal production cross section, which is determined
  by the charged scalar mass (see Fig.~\ref{lo_cros}).
  To study the impact of other model parameters, results presented in
  Fig.~\ref{fig:resultsFastSimAll} were scaled to 
  the signal production cross section times the branching ratio in the semi-leptonic channel of 1\,fb.
  Also, for comparison of the experimental sensitivity at \secondstage and
  \thirdstage, the same integrated luminosity of 4\,ab$^{-1}$ was assumed
  for both CLIC high energy running stages.
  Scaled significance values, after applying correction of Eq.(\ref{correction}), are presented 
  in Figure~\ref{fig:resultsFastSimScaled}.
  Results show that the best signal--background separation can be
  obtained for scenarios with scalar mass difference
  20\,GeV $<$ \chmassdiff $<$ 50\,GeV.
  For smaller mas differences, the impact of overlay events limits the
  experimental sensitivity while for higher mass differences,
  backgrounds dominated by processes with real $W^\pm$ production are
  more difficult to suppress.
  For a given production cross section, there is no additional systematic
  dependence of the sensitivity on the charged scalar mass,
  $m_{H^\pm}$.
  The sensitivity to charged IDM scalar production at \secondstage CLIC
  seems to be slightly better than at \thirdstage, which is most
  likely due to the smaller boost of the produced IDM scalars and $W$ bosons,
  resulting in better reconstruction of the di-jet invariant mass
  (see Fig.~\ref{fig:distributions}).

\section{Summary}

Prospects for observing production of new heavy scalar particles
were studied for CLIC high energy running stages.
Pair-production of heavy charged scalars was studied 
in the framework of the Inert Doublet Model, based on the benchmark scenarios proposed in \cite{Kalinowski:2018ylg}.
The expected final state, resulting from the charged scalar decays,
consists of two (real or virtual) W$^\pm$ bosons and a large missing
energy-momentum from the two escaping neutral scalars H, dark matter
candidates. 
For low mass scenarios and collision energies up to 500\,GeV, production rates are high and the di-lepton final state can be used as a discovery channel \cite{Zarnecki:2020swm}. 
However, this is no longer the case for scalar masses accessible only at higher collision energies.
For high mass scenarios considered in this study, the semi-leptonic final state was considered,
offering higher decay rates and hence also higher statistical significance than the
leptonic channels studied previously.
The CLIC potential was studied for 
5 selected IDM scenarios with the full detector simulation based on \geant
and for a complete set of 23 high mass IDM benchmark scenarios using the \delphes fast
simulation framework.
The CLICdet model for \delphes was extended to take into account the \ggtohad overlay
events.
This beam-induced background is crucial for the analysis, in
particular for signal scenarios with low scalar mass differences, when
the virtual W boson decay products are very soft.
Results of the study indicate that charged IDM scalars with masses up
to 1\,TeV can be detected at high energy running stages of CLIC.
For low scalar masses, the expected significance of the observation reaches
levels of up to about 40 standard deviations. 
Presented results indicate that semi-leptonic final state can be an important discovery channel also at other high energy lepton colliders, as 1\,TeV ILC \cite{bambade2019international} or Muon Collider \cite{Delahaye:2019omf}.
\section*{Acknowledgements}

The work was carried out in the framework of the CLIC detector and
physics (CLICdp) collaboration.
We thank collaboration members for fruitful discussions, valuable
comments and suggestions.
This work benefited from services provided by the ILC Virtual
Organisation, supported by the national resource providers of the EGI
Federation. This research was done using resources provided by the
Open Science Grid, which is supported by the National Science
Foundation and the U.S. Department of Energy's Office of Science.
The work was partially supported by the National Science Centre
(Poland) under OPUS research project no. 2017/25/B/ST2/00496
(2018-2021).

\printbibliography[title=References]



\end{document}